\documentclass[authorversion, sigconf]{acmart}

\usepackage{balance}

\def\BibTeX{{\rm B\kern-.05em{\sc i\kern-.025em b}\kern-.08emT\kern-.1667em\lower.7ex\hbox{E}\kern-.125emX}}
    
\copyrightyear{2019}
\acmYear{2019}
\setcopyright{acmcopyright}
\acmConference[CCS '19] {2019 ACM SIGSAC Conference on Computer and Communications Security}{November 11--15, 2019}{London, United Kingdom}
\acmPrice{15.00}
\acmDOI{10.1145/3319535.3339814}
\acmISBN{978-1-4503-6747-9/19/11} 

\usepackage{booktabs}

\usepackage{framed}

\usepackage{multirow}

\usepackage{subcaption}
\usepackage{multirow}

\usepackage[ruled,noend]{algorithm2e}

\usepackage{color}

\usepackage{enumitem}

\usepackage{pifont}
\newcommand{\cmark}{\ding{51}}%
\newcommand{\xmark}{\ding{55}}%

\usepackage{makecell}
\newcommand{\tabitem}{~~\llap{-}~~}

\newcommand{\system}{LightBox}

\newcommand{\inlsec}[2][0.25]{\vspace{#1\baselineskip}\noindent\textbf{#2}. }

\newcommand{\algstate}[1]{\texttt{#1}}
\newcommand{\us}{$\mu$s}

\newcommand{\moresp}[1]{\vspace{#1\baselineskip}}

\begin{document}

\fancyhead{}

\title{LightBox: Full-stack Protected Stateful Middlebox \\at Lightning Speed}

\author{Huayi Duan$^1$, Cong Wang$^1$, Xingliang Yuan$^2$, Yajin Zhou$^{34}$, Qian Wang$^5$, and Kui Ren$^{34}$}
\affiliation{%
  \institution{$^1$City University of Hong Kong and City University of Hong Kong Shenzhen Research Institute; $^2$Monash University; \\ $^3$College of Computer Science and Technology, School of Cyber Science and Technology, Zhejiang University;\\
  $^4$Alibaba-Zhejiang University Joint Research Institute of Frontier Technologies; \\
  $^5$School of Cyber Science and Engineering, Wuhan University;\\
  hy.duan@my.cityu.edu.hk, congwang@cityu.edu.hk, xingliang.yuan@monash.edu, \\
  yajin\_zhou@zju.edu.cn, qianwang@whu.edu.cn, kuiren@zju.edu.cn}
}

%
\begin{abstract}
Running off-site software middleboxes at third-party service providers has been a popular practice. However, routing large volumes of raw traffic, which may carry sensitive information, to a remote site for processing raises severe security concerns. Prior solutions often abstract away important factors pertinent to real-world deployment. In particular, they overlook the significance of metadata protection and stateful processing. Unprotected traffic metadata like low-level headers, size and count, can be exploited to learn supposedly encrypted application contents. Meanwhile, tracking the states of 100,000s of flows concurrently is often indispensable in production-level middleboxes deployed at real networks.

We present LightBox, the first system that can drive off-site middleboxes at near-native speed with stateful processing and the most comprehensive protection to date. Built upon commodity trusted hardware, Intel SGX, LightBox is the product of our systematic investigation of how to overcome the inherent limitations of secure enclaves using domain knowledge and customization. First, we introduce an elegant virtual network interface that allows convenient access to fully protected packets at line rate without leaving the enclave, as if from the trusted source network. Second, we provide complete flow state management for efficient stateful processing, by tailoring a set of data structures and algorithms optimized for the highly constrained enclave space. Extensive evaluations demonstrate that LightBox, with all security benefits, can achieve 10Gbps packet I/O, and that with case studies on three stateful middleboxes, it can operate at near-native speed.
\end{abstract}



\maketitle

\section{Introduction}
\label{sec:intro}
Middleboxes underpin the infrastructure of modern networks, where they undertake critical network functions for performance, connectivity, and security~\cite{Walfish2004}. Recently, a paradigm shift of migrating software middleboxes (aka virtual network functions) to professional service providers, e.g., public cloud, is taking place for the promising security, scalability and management benefits~\cite{benson2011, SherryHS12, gibb2012}. Its potential to enable a billion-dollar marketplace has already been widely recognized~\cite{zscaler_business}.

\begin{figure}
  \centering
  \includegraphics[width=.65\linewidth]{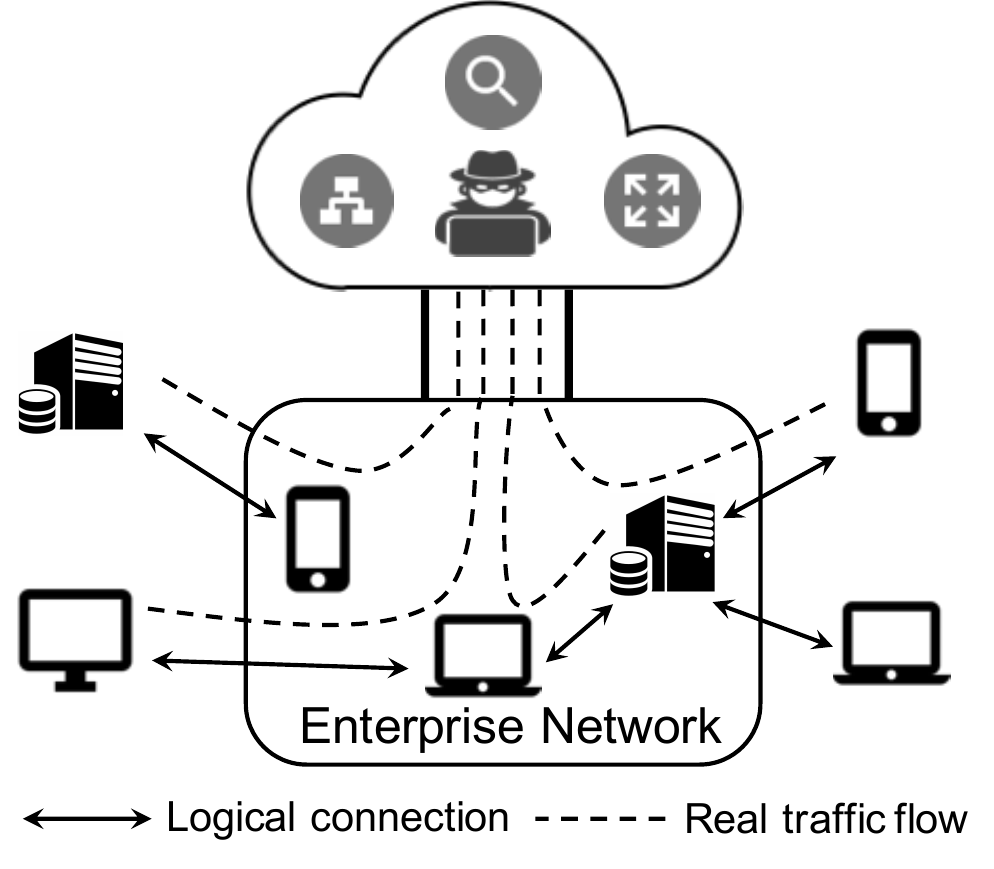}
  \caption{Large volumes of traffic are redirected to service providers for middlebox processing, creating a unique vantage point for adversaries to learn sensitive information.}
  \label{fig:influx}
\end{figure}

According to Zscaler~\cite{zscaler_data}, petabytes of traffic are now routed to its cloud-based security platform for middlebox processing every single day, and the number is still growing. Along with this seemingly unstoppable momentum comes an unprecedented security concern: how can end users be assured that their private information buried in the huge volumes of traffic, is not unauthorizedly leaked while being processed by the service provider (Fig.~\ref{fig:influx})? We are witnessing increasing and diversifying data breaches by service providers nowadays~\cite{data_breach}, yet embarrassingly we are also facing a daunting situation where full-scale traffic inspection seems mandatory to thwart stealthy threats~\cite{zscaler_ssl}. In light of this, a reassuring solution should be capable of protecting sensitive traffic while retaining necessary middlebox functionality.

Over the past few years, a number of approaches have been proposed to address the problem above, and they can be categorized into two broadly defined classes: \emph{software-centric} and \emph{hardware-assisted}. The first line of solutions~\cite{SherryCRS15,LanSP16,fan2017spabox,AsgharMSDK16,YuanWLW16,canard2017blindids} often rely on tailored cryptographic schemes. They are advantageous in providing provable security without hardware assumption, but often limited in functionality and sometimes in performance. The second line of solutions move middleboxes into a trusted execution environment, mostly Intel SGX enclave~\cite{mckeen2013}. Hardware-assisted designs provide generally better functionality and performance than software-centric approaches. In this regard, efforts with particular focus on the modular design and programmability of secure 
middleboxes~\cite{coughlin2017,han2017,trach2018shieldbox,poddar2018safebricks}, deployment consideration~\cite{naylor2017and}, and code protection~\cite{poddar2018safebricks} have been actively made.

As with these designs, we prioritize the consideration of handling the intrinsic complexity and stringent performance requirements of middleboxes, and leverage the SGX enclave as a starting point to develop a secure middlebox system. We observe that while previous solutions have claimed the benefits and practicality of SGX-enabled design, they largely overlooked several key factors that are highly pertinent to the real-world deployment of off-site middleboxes. We have identified two important aspects and provided encouraging results, which we hope will lead the secure middlebox-as-a-service to the practical realm and stimulate its massive adoption.

\subsection{Motivation}
\inlsec[0]{Necessity for metadata protection} 
Most existing designs only consider the protection of application payloads while redirecting traffic to remote middleboxes. Few of them protect traffic metadata, including \emph{low-level packet headers, packet size, count and timestamps}. Such metadata is information-rich and highly exploitable. The importance of hiding communication metadata (e.g., IP addresses) has been reiterated in recent years~\cite{van2015vuvuzela,Tyagi2017}. In the often cited example of Snowden Leaks, it is frustratedly put as \emph{``if you have enough metadata, you don't really need content''}~\cite{rusbridger2013}.

By just exploiting the seemingly innocent \emph{packet size, count and timing}, a variety of sophisticated traffic analysis attacks have been demonstrated: they can extract supposedly encrypted application contents such as website objects~\cite{wang2014effective}, VoIP conversations~\cite{WhiteMSM2011}, streaming videos~\cite{ReedK2017}, instant messages~\cite{Coull2014}, and user activities~\cite{conti2016analyzing}, by analyzing the distributions of metadata. 

Such metadata may be obtained by an adversary who can sniff traffic anywhere on the transmission path. In fact, aggregating tremendous user traffic to the middlebox service provider creates a unique vantage point for traffic analysis, because of the much enlarged datasets for correlating information and statistical inference. We should thus protect not only application payloads but all aforementioned traffic metadata --- what we dub \emph{full-stack protection}.

\begin{table}
\centering
\caption{Functionality and security characterization of representative solutions for secure middleboxes.}
\label{tbl:charaterization}
\resizebox{.48\textwidth}{!}{%
\begin{tabular}{@{}ccccccccc@{}}
\toprule & \multicolumn{3}{c}{Function} & \multicolumn{5}{c}{Protection} \\ 
\cmidrule(l){2-9} & Field & Op. & \multicolumn{1}{c|}{Stateful} & Meta & HDR & P/L & Rule & State       \\ \midrule
{\underline{\textit{Software-centric}}} & & & & & & & \\
\multicolumn{1}{c|}{BlindBox~\cite{SherryCRS15}} & P & PM & \xmark & \xmark & \xmark & \cmark & \cmark & N/A \\
\multicolumn{1}{c|}{YWLW16~\cite{YuanWLW16}} & P & PM & \xmark & \xmark & \xmark & \cmark & \cmark & N/A \\
\multicolumn{1}{c|}{SplitBox~\cite{AsgharMSDK16}} & H+P & RM & \xmark & \xmark & \cmark & \cmark & \cmark & N/A \\
\multicolumn{1}{c|}{BlindIDS~\cite{canard2017blindids}} & P & PM & \xmark & \xmark & \xmark & \cmark & \cmark & N/A \\
\multicolumn{1}{c|}{SPABox~\cite{fan2017spabox}} & P & REX & \xmark & \xmark & \xmark & \cmark & \cmark & N/A \\
\multicolumn{1}{c|}{Embark~\cite{LanSP16}} & H+P & RM & \cmark$^\ast$ & \xmark & \cmark & \cmark & \cmark & \cmark$^\ast$ \\ \midrule
{\underline{\textit{Hardware-assisted}}} & & & & & & & \\
\multicolumn{1}{c|}{S-NFV~\cite{shih2016}} & N/A & N/A & N/A & \xmark & \xmark & \xmark & \xmark & \cmark \\
\multicolumn{1}{c|}{TrustedClick~\cite{coughlin2017}} & H+P & GN & \xmark & \xmark & \xmark & \cmark & \cmark & N/A \\
\multicolumn{1}{c|}{SGX-BOX~\cite{han2017}} & P & GN & \cmark & \xmark & \xmark & \cmark & \cmark & \xmark \\
\multicolumn{1}{c|}{mbTLS~\cite{naylor2017and}} & P & GN & N/S & \xmark & \xmark & \cmark & \cmark & N/S \\
\multicolumn{1}{c|}{ShieldBox~\cite{trach2018shieldbox}} & H+P & GN & \xmark & \xmark & \xmark & \cmark & \cmark & N/A \\
\multicolumn{1}{c|}{SafeBrick~\cite{poddar2018safebricks}} & H+P & GN & \cmark$^\ast$ & \xmark & \cmark & \cmark & \cmark & \cmark$^\ast$ \\
\multicolumn{1}{c|}{\textbf{LightBox}} & H+P & GN & \cmark & \cmark & \cmark & \cmark & \cmark & \cmark \\
\bottomrule
\multicolumn{9}{l}{
\makecell[l]{
\underline{Notations}\\
\tabitem Field: which fields are processed, H (L2-L4 headers) and/or P (L4 payload). \\
\tabitem Op. (operation): PM (exact string pattern matching) $\subset$ RM (range matching) \\ $\qquad\qquad\qquad\quad$ $\subset$ REX (regular expression matching) $\subset$ GN (generic functions).\\
\tabitem Stateful: whether generic \emph{flow-level} stateful processing is supported. \\
\tabitem Meta: packet size, count and timestamp.\\
\tabitem HDR: L2-L4 headers, e.g., ip address, port number, and TCP/IP flags. \\
\tabitem P/L: L4 payload, including all application content. \\
\tabitem Rule: middlebox processing rules, e.g., attacking signatures. \\
\tabitem State: \emph{flow-level} states, e.g., connection status, statistics and stream buffers. \\
\tabitem N/A: the feature is not considered by design. \\
\tabitem N/S: the feature may be potentially supported, but not explicitly described. \\
~~\llap{$\ast$}~~: Embarks considers an ad hoc web proxy, but not generic stateful processing. \\
~~\llap{$\ast$}~~: SafeBrick considers a simple stateful firewall only.
}
}
\end{tabular}
}
\end{table}

\inlsec{Necessity for stateful middlebox}
In contrast to L2 switches and L3 routers that process each packet independently, advanced middleboxes need to track various flow-level states to implement complex functionality~\cite{jamshed17mos}. For example, intrusion detection systems typically keep per-flow stream buffers to eradicate cross-packet attack patterns~\cite{jamshed2012kargus,Snort}; proxies and load balancers maintain front/back-end connection states and packet pools to ensure end-to-end connectivity~\cite{HAProxy,Patel2013}. Thus, supporting stateful processing is an essential functionality requirement in realistic middlebox products.

However, even with the power of trusted hardware, it is technically challenging to develop a secure yet efficient solution due to the unique features of stateful middleboxes. In particular, the per-flow states range from a few hundreds of bytes~\cite{khalid2016} to multiple kilobytes~\cite{jamshed17mos}, and they need to stay tracked throughout the lifetime of flows or some expiration period. 
Moreover, it is not uncommon for production-level middleboxes to handle hundreds of thousands (or even more) of flows concurrently in real networks~\cite{de2014beyond, jamshed17mos, jamshed2012kargus, eisenbud2016maglev}. The resulting gigabytes of runtime memory footprint are not easily manageable by any secure enclaves.
Meanwhile, modern middleboxes feature packet processing delay within a few tens of microseconds~\cite{vIDS_CCS18,jamshed2012kargus,panda2016netbricks} --- a performance baseline that should be respected even if strong security guarantees are favored.

\inlsec{Characterization of prior arts}
We summarize existing solutions and compile their functionality and security features in Table~\ref{tbl:charaterization}. We define ``metadata'' as the L2-L4 headers, packet size, count and timestamps. For better characterization, we separately label the headers as ``HDR'', and the latter three as ``Meta''. Some solutions~\cite{SherryCRS15,shih2016,naylor2017and} do not target the outsourcing scenario and have different security goals, but we include them anyway for completeness.

Regarding security, none of the existing systems considers full-stack protection. The closest to us are Embark~\cite{LanSP16}, which applies deterministic encryption to each packet, and SafeBricks~\cite{poddar2018safebricks}, which uses L2 secure tunneling to forward packets to the enclave. Both of them encrypt packets individually and thus may be vulnerable to traffic analysis attacks exploiting packet size and count.

Regarding stateful processing, SGX-BOX~\cite{han2017} allows inspection on traffic streams that are reassembled outside enclave, and SafeBricks implements a simple stateful firewall for testing. But none of them considers the high flow concurrency exhibited in real networks and the challenges therein. Embark supports an ad hoc web proxy that caches static HTTP contents, but it cannot support generic stateful functions with arbitrary operations over flow state. We defer more discussions on related work to Section~\ref{sec:related}.

Note that we do recognize the meaningful explorations and contributions made by these prior arts with different focuses. We believe that our efforts will make it more convincing and confident to deploy and operate secure middleboxes in practical settings.

\subsection{Our Contribution}
We design and build \system, the first SGX-enabled secure middlebox system that can drive off-site middleboxes at near-native speed with stateful processing and full-stack protection. By systematically tackling many well-known limitations of SGX, from the lack of system services including network I/O, trusted timing and synchronization, to the unacceptable overhead of secure memory oversubscription, we have provided affirmative and satisfactory answers to the following two major research questions.

\moresp{0.5}
\emph{1. How to securely forward raw packets to a remote enclave, without leaking their low-level metadata, while still making them conveniently accessible at line rate?} 

In pursuit of this goal and in reminiscent of the classic kernel-driven \texttt{tun/tap} tunnel devices, we have developed a virtual network interface (VIF) called \texttt{etap} (“enclave tap”). It allows access to packets without leaving the enclave, as if from the source network where the packets originate.
For full packet protection, our design guarantees that the raw packets with L2-L4 headers are entirely delivered via a secure tunnel (e.g., TLS) terminated at a trusted enterprise gateway and the enclave. To frustrate traffic analysis, we pack the raw packets in a back-to-back manner and transmit them as continuous application payloads. As a result, the packet boundaries are obscured in the encrypted stream, so are the packet size and count. As a by-product of \texttt{etap}, we create a trusted clock for high-resolution and reliable timing inside the enclave.

We show how to progressively optimize the performance of \texttt{etap} with lock-free rings, cache line protection, and disciplined batching, so that it can catch up with the rate of physical network interfaces. 

We also endeavor to improve the usability of \texttt{etap}, by further porting three networking frameworks on top of it: 1) an adaption layer of \texttt{libpcap}~\cite{libpcap}, 2) a lightweight TCP stream reassembly library~\cite{libntoh}, and 3) an advanced flow monitoring stack mOS~\cite{jamshed17mos}. These system efforts allow us to port or develop middleboxes that enjoy the security and performance benefits of \texttt{etap}, with little code modification. For example, a developer can write a middlebox in the mOS framework as usual, and then automatically turn it into a \system{} instance even without the knowledge of SGX.

\moresp{0.5}
\emph{2. How to enable the resource-demanding stateful middlebox processing within the highly constrained enclave space, without incurring unreasonably high overhead?}

As mentioned before, stateful middleboxes in realistic settings have a large memory demand, which is at odds with the limited enclave memory supply. The naive paging approach for oversubscribing enclave memory incurs substantial overhead~\cite{arnautov2016}, which is intolerable to middleboxes as confirmed by our experiments.
To reduce the enclave footprint, we propose to maintain only a small working set of states in the enclave, while keeping the vast remainder of them encrypted in untrusted memory, at the granularity of flow. While this general idea may seem natural at first sight, we show that the performance bottleneck can only be surmounted with carefully crafted data structures and algorithms.

Specifically, given the limited enclave resources, the data structures used to hold flow states must be very compact, and support efficient lookup, relocation, swapping and deletion of data items. To meet these requirements, we design a set of interlinked data structures. They separate the indexing and storage of flow states, enabling flexible lookup strategy. They also allow fast relocation of states inside enclave and swapping of states across enclave boundary, with very cheap pointer operations. Furthermore, after identifying the lookup procedure as the main roadblock on the critical path, we opt to the space-efficient cuckoo hashing for indexing flow states, and introduce a cache-friendly lookup algorithm to counteract the cache inefficiency of hashing-based scheme.

Efficiency aside, our design ensures the confidentiality, integrity, and freshness of the states throughout the management procedures.

\inlsec{Experiment}
Extensive evaluations show that with our optimized designs, \texttt{etap} allows in-enclave packet I/O at $10$Gpbs rate with full-stack protection. We instantiate \system~for three stateful middleboxes, which are arguably more complicated than any of those tested by prior arts, and evaluate each of them against the native version and a strawman variant solely relying on EPC paging. \system~incurs negligible packet delay inflation to the native processing for the two middleboxes with $0.5$KB and $5.5$KB per-flow state, and moderate delay to the most complicated middlebox with $11.4$KB per-flow state. It maintains constant performance and achieves multi-factor speedup over the strawman when tracking $100,000$s of flows. The performance gap is widened as more flows are tracked and more severe paging penalty is imposed on the latter. Last but not the least, \system~ can achieve native speed on a real CAIDA trace for two of the three middleboxes under testing, and $2\times$ speedup over the strawman for the other unoptimized one.

\inlsec{Lessons learned}
From our experience in designing, building and evaluating \system, we have indeed challenged the common perception that one can run applications in secure enclaves, SGX in particular, at native speed almost without technical efforts but mundane code porting. Our results unveil that for realistic workloads that are security-critical and performance-sensitive, domain-specific design and optimization become a must to bypass the intricacies of secure enclaves. We hope to raise practitioners' awareness that, as a particular lesson, the proper use of memory-efficient and cache-friendly data structures and algorithms will make a big difference to the performance of those enclave applications.

\section{Overview}
\label{sec:overview}
\subsection{Service Model}
\label{subsec:service_model}
In a realistic service model, the enterprise redirects its traffic to the off-site middlebox hosted by the service provider for processing~\cite{zscaler_cloud}.
We assume that the middlebox code is not necessarily private and may be known to the service provider. This matches practical use cases where the source code is free to use, but only bespoke rule sets~\cite{snort_rulesub} are proprietary. We also consider a single middlebox.
These simplifications allow us to concentrate on presenting the core designs of \system. Nonetheless, we stress that \system\ can be readily adapted to support service function chaining~\cite{sun2017nfp} and disjoint service providers~\cite{poddar2018safebricks}, which mostly involves only changes to the service launching phase. We postpone the discussions of other service models to Appendix \ref{sec:extension}.

\inlsec{Traffic forwarding}
For ease of exposition, we consider the bounce model with one gateway~\cite{SherryHS12,LanSP16}: both inbound and outbound traffic is redirected from an enterprise gateway to the remote middlebox for processing and then bounced back. The other direct model, where traffic is routed from the source network to the remote middlebox and then directly to the next trusted hop, i.e., the gateway in the destination network~\cite{poddar2018safebricks}, can be trivially supported by installing one \texttt{etap-cli} (see Section~\ref{subsec:architecture}) on each gateway. 

The communication endpoints themselves (e.g., a client in the enterprise network and an external server) may transmit data via a secure connection. To enable such already encrypted traffic to be processed by the middlebox, the gateway needs to intercept the secure connection and decrypt the traffic before redirection. We follow the common practice~\cite{poddar2018safebricks,naylor2017and,han2017} to handle this issue. In particular, the gateway will receive the session keys from the endpoints to perform the interception, unbeknownst to the middlebox.

A dedicated high-speed connection will be typically established for traffic redirection~\cite{SherryHS12}. Such services have been widely provisioned nowadays, for example AWS Direct Connect~\cite{aws_directconnect}, Azure ExpressRoute~\cite{azure_expressroute}, and Google Dedicated Interconnect~\cite{google_dedicatedinterconnect}. The off-site middlebox, while being secured, should also be able to process packet at line rate to benefit from such dedicated links.

\begin{figure}
  \centering
  \includegraphics[width=0.9\linewidth]{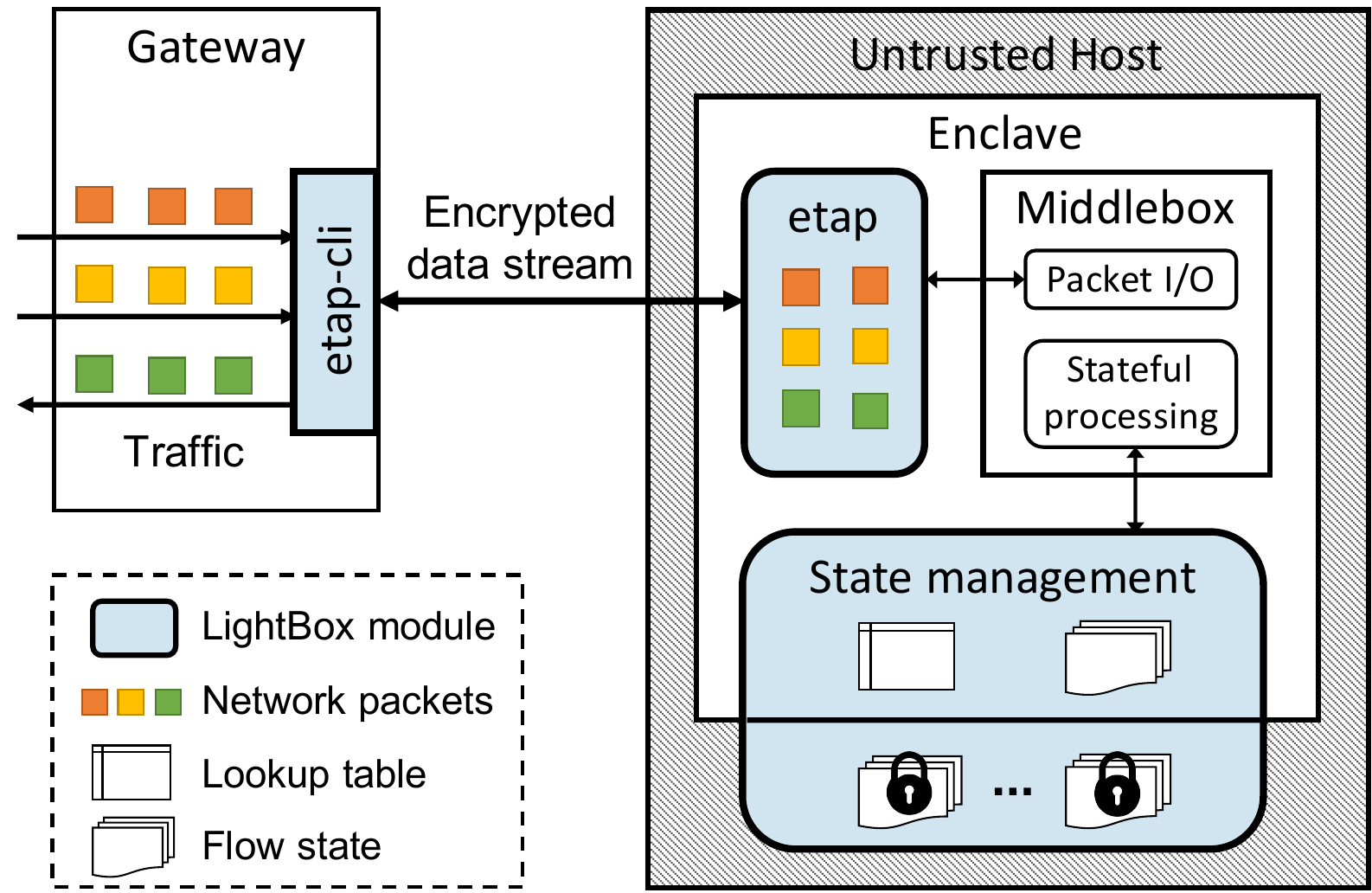}
  \caption{Overview of \system\ components.}
  \label{fig:system_arch}
\end{figure}

\subsection{SGX Background} 
\label{app:sgx}
SGX introduces a trusted execution environment called enclave to shield code and data with on-chip security engines. It stands out for the capability to run generic code at processor speed, with practically strong protection. Despite the benefits, it has several limitations. First, common system services cannot be directly used inside the enclave. Access to them requires expensive context switching to exit the enclave, typically via a secure API called \texttt{OCALL}. Second, memory access in the enclave incurs performance overhead. The protected memory region used by the enclave is called Enclave Page Cache (EPC). It has a conservative limit of $128$MB in current product lines. Excessive memory usage in the enclave will trigger EPC paging, which can induce prohibitive performance penalties~\cite{arnautov2016}. Besides, the cost of cache miss while accessing EPC is higher than normal, due to the cryptographic operations involved during data transferring between CPU cache and EPC. While such overhead may be negligible to certain applications, it becomes crucial to middleboxes with stringent performance requirements.

\subsection{\system\ Overview}
\label{subsec:architecture}

\inlsec[0]{\system\ components}
\system\ leverages an SGX enclave to shield the off-site middlebox. 
As shown in Fig.~\ref{fig:system_arch}, a LightBox instance comprises two modules in addition to the middlebox itself: a virtual network interface \texttt{etap} and a state management module. The former is semantically equivalent to a physical network interface card (NIC), allowing packets I/O at line rate within the enclave. The latter provides automatic and efficient memory management of the large amount of flow states tracked by the middlebox.

The \texttt{etap} device is peered with one \texttt{etap-cli} program installed at the enterprise gateway. We establish a persistent secure channel between the two to tunnel the raw traffic, which is transparently encoded/decoded and encrypted/decrypted by \texttt{etap}. The middlebox and upper networking layers can directly access raw packets via \texttt{etap} without leaving the enclave.

The state management module maintains a small flow cache in the enclave, a large encrypted flow store in the untrusted memory, and an efficient lookup data structure in the enclave. The middlebox can lookup or remove state entries by providing flow identifiers. In case a state is not present in the cache but in the store, the module will automatically swap it with a cached entry.

\inlsec{Secure service launching}
The enterprise needs to attest the integrity of the remotely deployed \system\ instance before launching the service. This is realized by the standard SGX attestation utility~\cite{anati2013}. Specifically, the enterprise administrator can request a security measurement of the enclave signed by the CPU, and interact with Intel's IAS API for verification. During attestation, a secure channel is established to pass configurations, e.g., middlebox processing rules, \texttt{etap} ring size and flow cache size, to the \system\ instance. Due to the space limit, we skip the verbose description here. We remark that for the considered service scenario where only two parties (the enterprise and the server provider) are involved, a basic attestation protocol between the two and Intel IAS is sufficient. 

\subsection{Adversary Model}
\label{subsec:threat_model}
In line with SGX's security guarantee, we consider a powerful adversary. We assume that the adversary can gain full control over all user programs, OS and hypervisor, as well as all hardware components in the machine, with the exception of processor package and memory bus. It can obtain a complete memory trace for any process, except those running in the enclave. It is also capable of observing network communications, modifying and dropping packets at will. In particular, the adversary can log all network traffic and conduct sophisticated inference to mine useful information. 
Our goal here is to thwart practical traffic analysis attacks targeting the \emph{original packets} that are intended for processing at the off-site middleboxes.

Like many SGX applications~\cite{arnautov2016,schuster2015,hunt2016,orenbach2017,Mishra2018}, we consider side-channel attacks~\cite{moghimi2017cachezoom,gotzfried2017cache,xu2015,shinde2016prevent,weichbrodt2016} out of scope. They can be orthogonally handled by corresponding countermeasures~\cite{shinde2016prevent,costan2016,chen2017detecting,seo2017,Gruss2017}. That said, we fully recognize the security benefits and limitations of SGX, and understand that they are still under rapid iteration~\cite{wang2017leaky}. It is yet by far, arguably, the most practical solution for a wide class of complicated tasks including secure middleboxes. 

We do not deal with denial-of-service attacks. The middlebox code is assumed to be correct. Also, we assume that the enterprise gateway is always trusted and it does not have to be SGX-enabled. 

\section{The \texttt{etap} Device}
\label{sec:etap}
The ultimate goal of \texttt{etap} is to enable in-enclave access to the packets intended for middlebox processing, \emph{as if they were locally accessed from the trusted enterprise networks.} Towards this goal, we set forth the following design requirements.
\begin{itemize}
\item \emph{Full-stack protection}: when the packets are transmitted in the untrusted networks, and when they traverse through the untrusted platform of the service provider, none of their metadata as defined in Section~\ref{sec:intro} is directly leaked.
\item \emph{Line-rate packet I/O}: \texttt{etaps} should deliver packets at a rate that can catch up with a physical network interface card (NIC), without capping the middlebox performance. A pragmatic performance target to shoot is $10$Gbps. 
\item \emph{High usability}: to use \texttt{etap}, we need to impose as few changes as possible to the secured middlebox. This implies that if certain network frameworks are used by the middlebox, they should be seamlessly usable inside the enclave too.
\end{itemize}

\begin{figure}
  \centering
  \includegraphics[width=.95\linewidth]{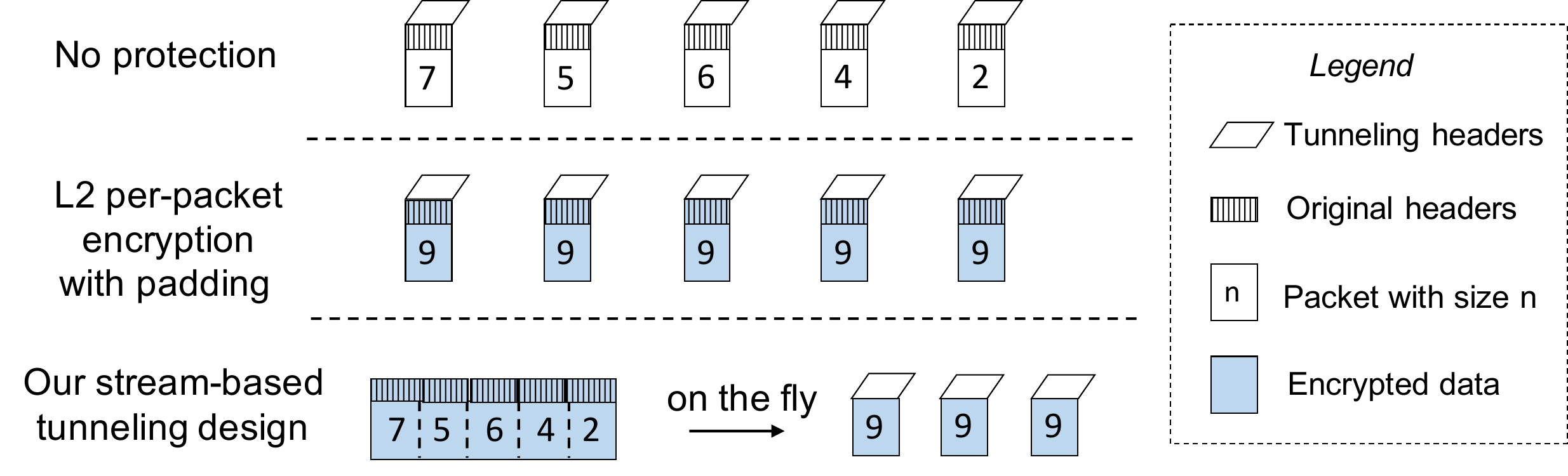}
  \caption{Illustration of secure tunneling options.}
  \label{fig:tun_ill}
\end{figure}

\subsection{Overview}
\label{subsec:etap_rationale}
To achieve full-stack protection, an intuitive idea would be to securely \emph{tunnel} the packets between the gateway and the enclave: the original packets are encapsulated and encrypted as the payloads of new packets, which contain non-sensitive header information (i.e., the IP addresses of the gateway and the middlebox server).

The naive way of encapsulating and encrypting packets individually, as used in L2 tunneling solution like IPSec, does not suffice for our purpose --- it does not protect information pertaining to individual packets, including size, timestamp, and as a result, packet count. Padding each packet to the maximum size may hide exact packet size, but it incurs unnecessary bandwidth inflation, and still cannot hide the count and timestamps.

We thus consider encoding the packets as a continuous stream, which is treated as application payloads and transmitted via a secure channel (e.g., TLS). Such streaming design obfuscates packet boundaries, thus hiding the metadata we want to protect, as illustrated in Fig.~\ref{fig:tun_ill}. From a system perspective, the key to this approach is the VIF \texttt{tun/tap}\footnote{https://www.kernel.org/doc/Documentation/networking/tuntap.txt} that can be used as an ordinary NIC to access the tunneled packets, as widely adopted by popular products like OpenVPN. While there are many userspace TLS suites and some of them even have handy SGX ports~\cite{wolfssl,mbedtls,talos}, the \texttt{tun/tap} device itself is canonically driven by the untrusted OS kernel. That is, even if we can terminate the secure channel inside the enclave, the packets are still exposed when accessed via the untrusted \texttt{tun/tap} interface.

This inspires us to develop \texttt{etap} (the ``enclave \texttt{tap}''), which manages packets inside the enclave and enables direct access to them without exiting. The naming comes from the convention that \texttt{tap} is for L2 packets (more precisely, frames) while \texttt{tun} is for L3 packets, and we aim to protect the L2 header as well. From the middlebox's point of view, accessing packets in the enclave via \texttt{etap} is identical to accessing them via a real NIC in the local enterprise networks.

\begin{figure}
  \centering
  \includegraphics[width=.95\linewidth]{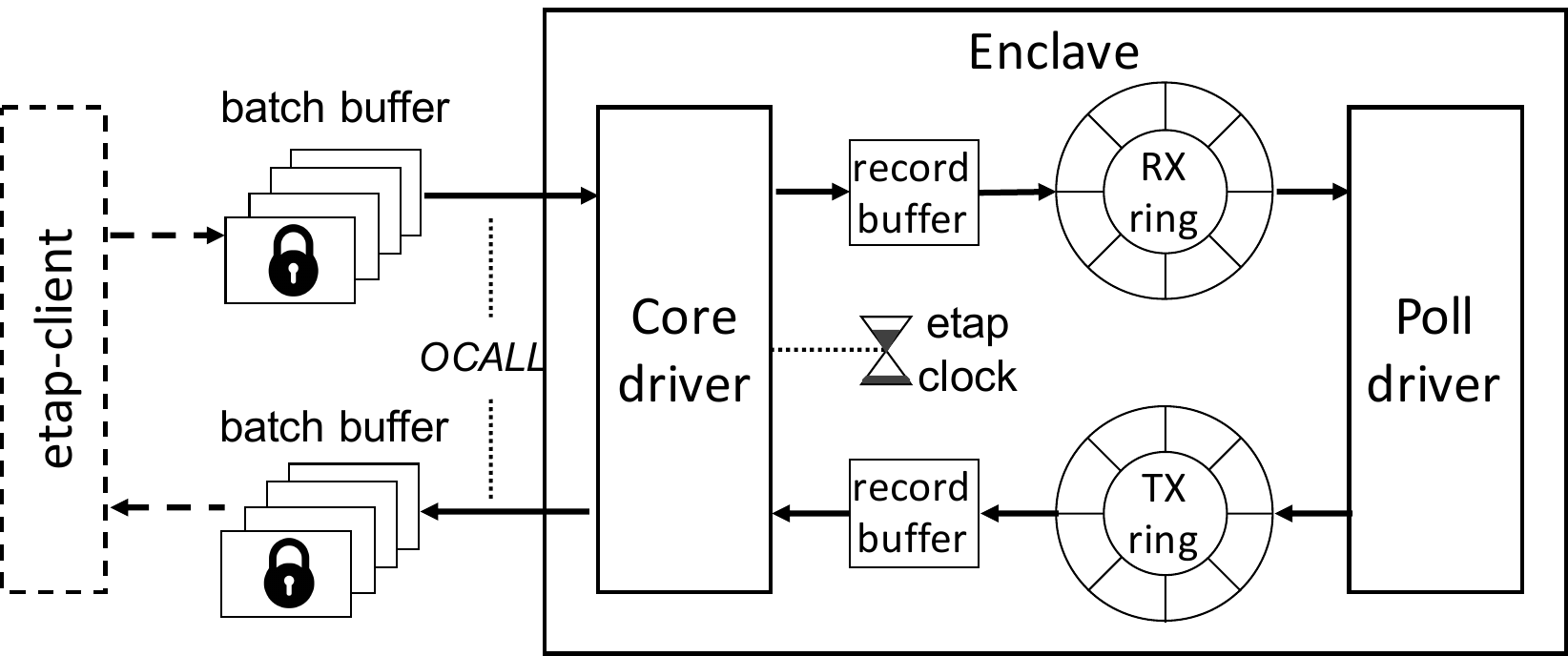}
  \caption{The architecture of \texttt{etap}.}
  \label{fig:etap_arch}
\end{figure}

\subsection{Architecture}
\label{subsec:etap_arch}
The major components of \texttt{etap} are depicted in Fig.~\ref{fig:etap_arch}. Every \texttt{etap} is peered with an \texttt{etap-cli} run by the gateway. The two share the same processing logic and since \texttt{etap-cli} operates as a normal program in the trusted gateway, we ignore a detailed description for it. A persistent connection will be established between the two for secure traffic tunneling. 
The \textrm{etap} peers will maintain a minimal traffic rate by injecting heartbeat packets to the tunnel.

At the heart of \texttt{etap} are two rings for queuing packet data: one for transmitting (TX) and the other for receiving (RX). A packet is described by a \texttt{pkt\_info} structure, which stores in order the packet size, timestamp, and a buffer for raw packet data. 
Two additional data structures are used in preparing and parsing packets: a record buffer holds decrypted data and some auxiliary fields inside the enclave; a batch buffer stores multiple records outside the enclave.

The \texttt{etap} device is powered by two drivers. The core driver coordinates networking, encoding and cryptographic operations; it also maintains a trusted clock to overcome the lack of high-resolution timing inside the enclave. The poll driver is used by middleboxes to access packets. The two drivers source and sink the two rings accordingly.
We discuss the support of multiple RX rings for multi-threaded middleboxes in Appendix~\ref{sec:multi_threading}.

\begin{figure}
  \centering
  \includegraphics[width=.9\linewidth]{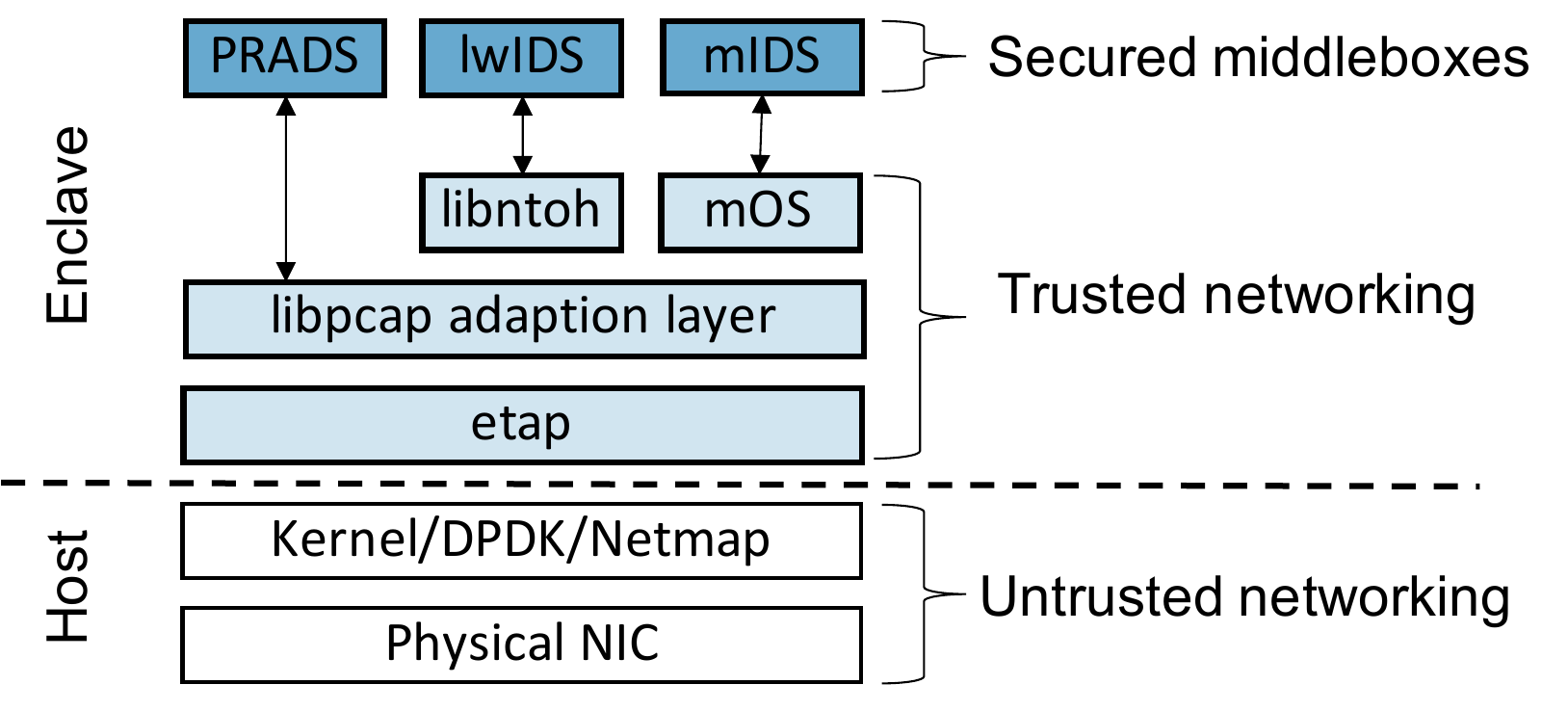}
  \caption{The networking stack enabled by \texttt{etap}.}
  \label{fig:etap_stack}
\end{figure}

\inlsec{Remark}
The design of \texttt{etap} is agnostic to how the \emph{real} networking outside the enclave is performed. 
For example, it can use standard kernel networking stack (this is currently used by us). For better efficiency, it can also use faster userspace networking frameworks based on DPDK~\cite{IntelDPDK} or netmap~\cite{Rizzo2012}, as shown in Fig.\ref{fig:etap_stack}.

\subsection{Drivers}
\label{subsec:etap_detail}

\begin{algorithm}[t]
\nl \label{algline:ocall}\algstate{ocall\_fill\_rx\_bat\_buf()}\;
\nl \label{algline:valid}\algstate{check\_memory\_safety(rx\_bat\_buf)}\;
\nl \label{algline:record_loop_beg}\ForEach{\algstate{\textup{enc\_rec in rx\_bat\_buf}}}{
\nl \algstate{rec\_buf = decrypt(enc\_rec)}\;
\nl \algstate{finish\_pending\_partial\_pkt(rec\_buf)}\;
\nl \label{algline:packet_loop_beg}\While{\algstate{\textup{has\_full\_pkt(rec\_buf)}}}{
    \nl \algstate{pkt\_info = parse\_next(rec\_buf)}\;
    \nl \label{algline:packet_loop_end}\algstate{push\_to\_rx\_etap\_ring(pkt\_info)}\;
}
\nl \label{algline:record_loop_end}\algstate{refresh\_pending\_partial\_pkt(rec\_buf)}
}
\caption{\texttt{etap} core driver's RX ring loop}
\label{alg:etap_core_driver}
\end{algorithm}

\inlsec[0]{Core driver}
\label{subsubsec:etap_core_driver}
Upon initialization, the core driver takes care of necessary handshakes (via \texttt{OCALL}) for establishing the secure channel and stores the session keys inside the enclave. 
The packets intended for processing are pushed into the established secure connection in a back-to-back manner, forming a data stream at the application layer. 
At the transportation layer they are effectively organized into contiguous records (e.g., TLS records) of fixed size (e.g., 16KB for TLS), which then at the network layer are broken down into packets of maximum size.
Each original packet is transmitted in the exact format of \texttt{pkt\_info}, so the receiver will be able to recover from the continuous stream the original packet by first extracting its length, the timestamp, and then the raw packet data. 
The core driver is run by its own thread. The pseudo code in Alg.~\ref{alg:etap_core_driver} outlines the the main RX loop. The TX side is similar and omitted here.

\inlsec{Trusted timing with \texttt{etap} clock}
\label{subsubsec:clock}
Middleboxes often demand reliable timing for packet timestamping, event scheduling, and performance monitoring. The timer should at least cope with the packet processing rate, i.e., at tens of microseconds. The SGX platform provides trusted relative time source~\cite{inteltrustedtime}, but its resolution is too low (at seconds) for our use case.
Alternative approaches resort to system time provided by OS~\cite{poddar2018safebricks} and on-NIC PTP clock~\cite{trach2018shieldbox}. 
Yet, they both access time from untrusted sources, thus subject to adversarial manipulation. 
Another system~\cite{zhang2016} fetches time from a remote trusted website, and its resolution (at hundreds of milliseconds) is still unsatisfactory for middlebox systems.

We design a reliable clock by taking \texttt{etap}'s architectural advantage. In particular, we treat \texttt{etap-cli} as a trusted time source to attach timestamps to the forwarded packets. The core driver can then maintain a clock (with proper offset) by updating it with the timetamp of each received packet. The resolution of the clock is determined by the packet rate, which in turn bounds the packet processing rate of the middlebox itself. Therefore, the clock should be sufficient for most timing tasks found in middleboxes.
Furthermore, we collate the clock periodically with the round-trip delay estimated by the moderately low-frequency heartbeat packets sent from \texttt{etap-cli}, in a way similar to the NTP protocol~\cite{ntp1985}. 
We stress that it is still an open problem to provide trusted and high-resolution time for SGX applications~\cite{inteltrustedtime,chen2017detecting}. The proposed \texttt{etap} clock fits well for middlebox processing in our targeted high-speed networks.

\inlsec{Poll driver}
The poll driver provides access to \texttt{etap} for upper layers. It supplies two basic operations, \texttt{read\_pkt} to pop packets from RX ring, and \texttt{write\_pkt} to push packets to TX ring.
Unlike the core driver, the poll driver is run by the middlebox thread. It has two operation modes. In the default blocking mode, a packet is guaranteed to be read from or write to \texttt{etap}: in case the RX (resp. TX) ring is empty (resp. full), the poll driver will spin until the ring is ready. In the non-blocking mode, the driver returns immediately if the rings are not ready. This will allow the middlebox more CPU time for other tasks, e.g., processing cached events.

\begin{figure}
    \centering
    \includegraphics[width=.9\linewidth]{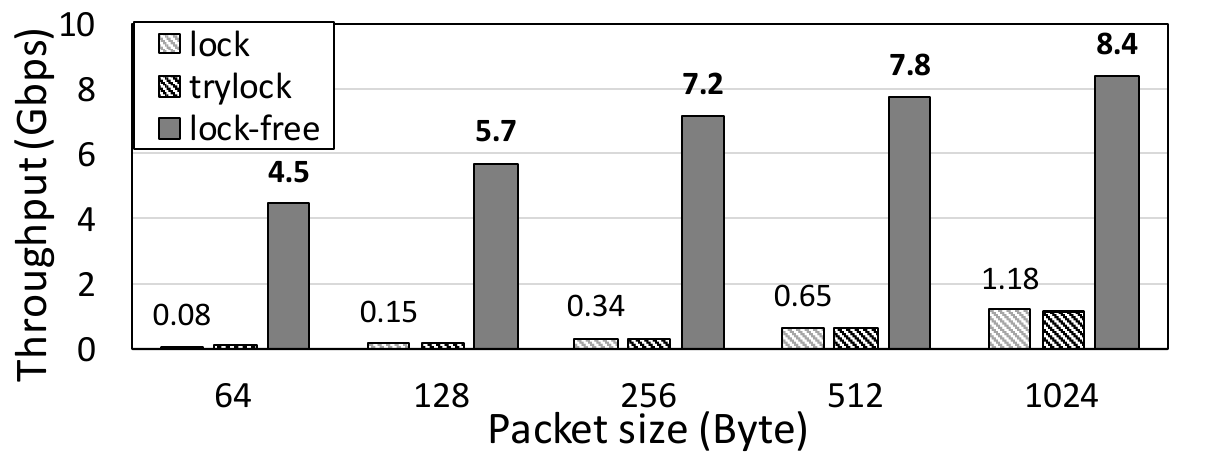}
    \caption{Performance of \texttt{etap} by applying three different synchronization mechanisms, without other optimizations suggested in Section~\ref{subsec:etap_perf}.}
    \label{fig:micro_etun_lock}
\end{figure}

\subsection{Security Analysis}
\label{subsec:etap_security}
The protection of application payloads in the traffic is obvious. We focus on the analysis of metadata. Also we discuss passive adversary only, because the active ones who attempt to modify any data will be detected by the employed authenticated encryption.

\inlsec{Metadata protection}
Imagine an adversary located at the ingress point of the service provider's network, or one that has gained full privilege in the middlebox server. She can sniff the entire tunneling traffic trace between the \texttt{etap} peers. As illustrated in Fig.~\ref{fig:tun_ill}, however, the adversary is not able to infer the \emph{packet boundaries} from the encrypted stream embodied as the encrypted payloads of observable packets, which have the maximum size most of the time. Therefore, she cannot learn the low-level headers, size and timestamps of the encapsulated individual packets in transmission. This also implies her inability to obtain the exact packet count, though this number is always bounded in a given period of time by the maximum and minimum possible packet size. 
Besides, the timestamp attached to the packets delivered by \texttt{etap} comes from the trusted clock, so it is invisible to the adversary.

As a result, a wide range of traffic analyses~\cite{wang2014effective,WhiteMSM2011,ReedK2017,Coull2014,conti2016analyzing} that directly leverage the metadata will be thwarted, as no such information is available to the adversary.

\inlsec{Beyond metadata}
Despite the aforementioned protection over metadata, we do not claim that our design defeats all possible inference attacks. For example, from the bursts in traffic the adversary may potentially learn the launching and termination of certain applications. It is also shown that some delicate analysis can glean meaningful information by looking at the total traffic volume and bursts~\cite{dyer2012peek}. A generic countermeasure would be to obfuscate exploitable traffic patterns by injecting calibrated noise~\cite{Wang2017WT}. We note that in-depth traffic analysis and mitigation is still a highly active research area~\cite{Imani2018ATW,8371242,Sirinam2018DFU}. In the context of middlebox outsourcing where an adversary may gain more visibility into the aggregated network traffic, our design can significantly raise the bar for realistic traffic analysis attacks.

\subsection{Performance Boosting}
\label{subsec:etap_perf}
While ensuring strong protection, \texttt{etap} is hardly useful if it cannot deliver packets at a practical rate. We therefore synergize several techniques to boost its performance. 

\inlsec{Lock-free ring}
The packet rings need to be synchronized between the two drivers of \texttt{etap}. We compare the performance of three approaches: a basic mutex (\texttt{sgx\_thread\_mutex\_lock}), a spinlock without context switching (\texttt{sgx\_thread\_mutex\_trylock}), and a classic single-producer-single-consumer lockless algorithm~\cite{lamport1977proving}. 
Our evaluation shows that the trusted synchronization primitives of SGX are too expensive for the use of \texttt{etap} (see Fig.~\ref{fig:micro_etun_lock}), so we base further optimizations on the lock-free design.

\inlsec{Cache-friendly ring access}
In the lock-free design, frequent updates on control variables will trigger a high cache miss rate, the penalty of which is amplified in the enclave.
We adapt the cache-line protection technique~\cite{lee2010lock} to our design to relieve this issue. It works by adding a set of new control variables local to the threads to reduce the contention on shared variables.
Our evaluations show that this optimization results in a performance gain up to $31\%$.

\inlsec{Disciplined record batching}
Recall that the core driver uses \texttt{bat\_buf} to buffer the records. The buffer size has to be properly set for best performance. If too small, the overhead of \texttt{OCALL} cannot be well amortized. If too large, the core driver needs longer time to perform I/O: this would waste CPU time not only for the core driver that waits for I/O outside the enclave, but also for a fast poll driver that can easily drain or fill the ring. Through extensive experiments, we find a batch size around $10$ to be a sweet spot that can deliver practically the best performance for different packet sizes in our settings (see Fig.~\ref{fig:net_io_batching}).

\subsection{Usability}
\label{subsec:etap_use}
A main thrust of \texttt{etap} is to provide convenient networking functions to in-enclave middleboxes, preferably without changing their legacy interfaces. On top of \texttt{etap}, we can port existing frameworks and build new ones. Here, we report our porting efforts of three of them, which greatly improve the usability of \texttt{etap}.

\inlsec{Compatibility with libpcap} 
Considering \texttt{libpcap} is widely used in networking frameworks and middleboxes for packet capturing~\cite{kohler2000click,Rizzo2012,libnids,PRADS,Snort}, we create an adaption layer that implements \texttt{libpcap} interfaces over \texttt{etap}, including the commonly used packet reading routines (e.g., \texttt{pcap\_loop}, \texttt{pcap\_next}), and filter routines (e.g., \texttt{pcap\_compile}). This layer allows many legacy systems, including the ones discussed in Section~\ref{sec:impl_inst}, to transparently access protected raw packets inside the enclave.

\inlsec{TCP reassembly}
This common function organizes the payloads of possibly out-of-order packets into streams for subsequent processing. To facilitate middleboxes demanding such functionality, we port a lightweight reassembly library \texttt{libntoh}~\cite{libntoh} on top of \texttt{etap}. It exposes a set of APIs to create stream buffers for new flows, add new TCP segments, and flush the buffers with callback functions. It is used in one of our middlebox case-studies. 

\inlsec{Advanced networking stack}
We also manage to port an advanced networking stack called mOS, which allows for programming stateful flow monitoring middleboxes~\cite{jamshed17mos}, into the enclave on top of \texttt{etap}. As a result, a middlebox built with mOS can automatically enjoy all security and performance benefits of \texttt{etap}, without the need for the middlebox developer to even have any knowledge of SGX. The porting is a non-trivial task as mOS has complicated TCP context and event handling, as well as more sophisticated payload reassembly logic than \texttt{libntoh}. Our current porting retains the core processing logic of mOS and only removes the threading features. 

Note that the two stateful frameworks above track flow states themselves, so running them inside the enclave efficiently requires delicate state management, as described in the next section.

\section{Flow State Management}
\label{sec:state_mgmt}
To get rid of the expensive application-agnostic EPC paging, a natural idea would be to carefully partition the working set of an SGX application into two parts, a small one that can fit in the enclave, and a large one that can securely reside in the untrusted main memory, while ensuring data swapping between the two in an on-demand manner. A related idea has been positively validated in a prior study~\cite{orenbach2017}. But it still falls within the paradigm of paging, and reports a slowdown over native processing of several factors, even for moderate working sets of a few hundreds of MBs. 

To effectively implement the idea above, we design a set of novel data structures specifically for managing flow states in stateful middleboxes. They are compact, such that collectively adding a few tens of MBs overhead to track one million flows concurrently. They are also interlinked, such that the data relocation and swapping involves only cheap pointer operations in addition to necessary data marshalling. To overcome the bottleneck of flow lookup, we further leverage the space-efficient cuckoo hashing to create a fast dual lookup algorithm. Altogether, our state management scheme introduces small and nearly constant computation cost to stateful middlebox processing, even with 100,000s of concurrent flows.

Note that we focus on flow-level states, which are the major culprits that overwhelm memory. Other runtime states, such as global counters and pattern matching engines, do not grow with the number of flows, so we leave them in the enclave and handled by EPC paging whenever necessary. Our experiments confirm that the memory explosion caused by flow states is the main source of performance overhead.

\begin{figure}
  \centering
  \includegraphics[width=.98\linewidth]{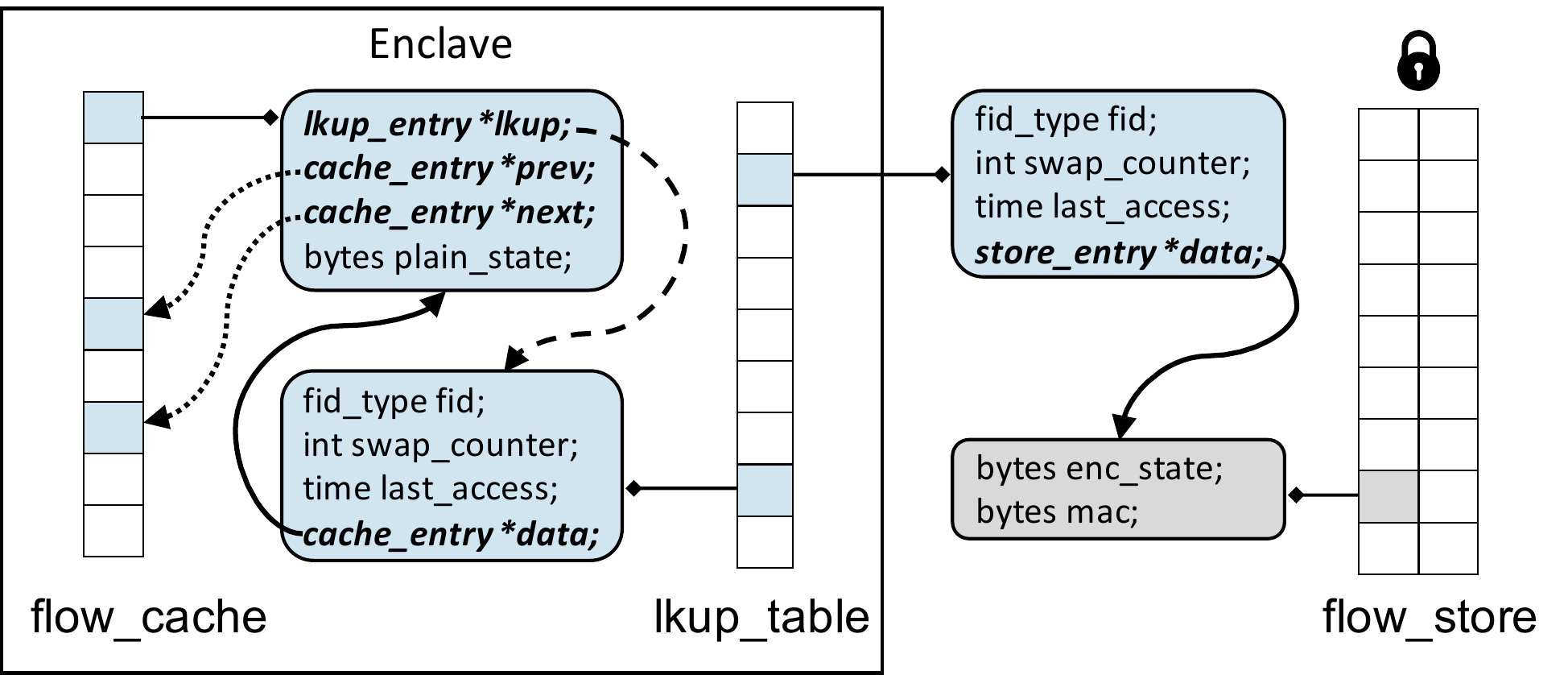}
  \caption{Data structures used in flow state management.}
  \label{fig:state_data_struct}
\end{figure}

\subsection{Data Structures}
\label{subsec:state_data}
The state management is centered around three abstract tables:
\begin{itemize}
\item \texttt{flow\_cache}, which maintains the states of a fixed number of active flows \emph{in the enclave};
\item \texttt{flow\_store}, which keeps the encrypted states of inactive flows \emph{in the untrusted memory};
\item \texttt{lkup\_table}, which allows fast lookup of all flow states from \emph{within the enclave}.
\end{itemize}
Among them, \texttt{flow\_cache} has a fixed capacity, while \texttt{flow\_store} and \texttt{lkup\_table} can grow as more flows are tracked. Our design principle is to keep the data structures of \texttt{flow\_cache} and \texttt{lkup\_table} functional and minimal, so that they can scale to millions of concurrent flows. Figure~\ref{fig:state_data_struct} gives an illustration of them.

\begin{itemize}
\item The \texttt{cache\_entry} holds raw state data. It keeps two pointers (dotted arrows) to implement the Least Recently Used (LRU) eviction policy, and links (dashed arrow) to a \texttt{lkup\_entry}.
\item The \texttt{store\_entry} holds encrypted state data and authentication MAC. It is maintained in untrusted memory so does not consume enclave resources.
\item The \texttt{lkup\_entry} stores \texttt{fid}, a pointer (solid arrow) to either \texttt{cache\_entry} or \texttt{store\_entry}, and two small fields. The \texttt{fid} represents the conventional 5-tuple to identify flows. The \texttt{swap\_count} serves as a monotonic counter to ensure the freshness of state; it is initialized to a random value and incremented by 1 on each encryption. The \texttt{last\_access} assists flow expiration checking, it is updated with \texttt{etap} clock on each flow tracking. Note that the design of \texttt{lkup\_entry} is independent of the underlying lookup structure, which for example can be plain arrays, search trees or hash tables.
\end{itemize}
  
The data structures above are succinct, making it efficient to handle high flow concurrency. Assume $8$B (byte) pointer and $13$B \texttt{fid}, then \texttt{cache\_entry} uses $24$B per \emph{cached flow} and \texttt{lkup\_entry} uses $33$B per \emph{tracked flow}. Assume $16$K cache entries and full utilization of the underlying lookup structure, then tracking $1$M flows requires only $33.8$MB enclave memory besides the state data itself.

\subsection{Management Procedures}
\label{subsec:state_mgmt}
We refer to \emph{flow tracking} as the process of finding the correct flow state on a given \texttt{fid}. It takes place in the early stage of the packet processing cycle. The identified state may be accessed anywhere and anytime afterwards~\cite{khalid2016,Kablan2017}. Thus, it should be pinned in the enclave immediately after flow tracking to avoid being accidentally paged out. The full flow tracking procedure is described in Alg.~\ref{alg:state}.

\inlsec{Initialization}
For efficiency, we preallocate entries for all three components. During initialization, a random key is generated and stored inside the enclave for the required authenticated encryption.

\inlsec{Flow tracking}
Given a \texttt{fid}, we first search through \texttt{lkup\_table} to check if the flow has been tracked. If it is found in \texttt{flow\_cache}, we will relocate it to the front of the cache by updating its logical position via the pointers, and return the raw state data. If it is found in \texttt{flow\_store}, we will swap it with the LRU victim in \texttt{flow\_cache}. In case of a new flow, an empty \texttt{store\_entry} is created for the swapping. 
The entry swapping involves a series of strictly defined operations: 1) Checking memory safety of the candidate \texttt{store\_entry}; 2) Encrypting the victim \texttt{cache\_entry}; 3) Decrypting the \texttt{store\_entry} to the just freed \texttt{flow\_cache} cell; 4) Restoring the lookup consistency in the \texttt{lkup\_entry}; 5) Moving the encrypted victim \texttt{cache\_entry} to \texttt{store\_entry}. At the end of flow tracking, the expected flow state will be cached in the enclave and returned to the middlebox.

\inlsec{Tracking termination}
The tracking of a flow can be explicitly terminated (e.g., upon seeing \texttt{FIN} or \texttt{RST} flag). When this happens, the corresponding \texttt{lkup\_entry} is removed and the \texttt{cache\_entry} is nullified. This will not affect \texttt{flow\_store}, as the flow has already been cached in the enclave.

\inlsec{Expiration checking}
We periodically purge expired flow states to avoid performance degradation. The last access time field will be updated at the end of flow tracking for each packet using the \texttt{etap} clock. The checking routine will walk through the \texttt{lookup\_table} and remove inactive entries in the tables.

\begin{algorithm}[t]
\KwIn{A \algstate{fid} extracted from input packet.}
\KwOut{The \texttt{state} of flow \texttt{fid}.}

\nl \algstate{entry = flow\_cache\_cuckoo\_lkup(fid)}\;
\nl \uIf(\tcp*[f]{\texttt{flow\_cache} miss}){\textup{\texttt{entry} empty}}{
\nl     \algstate{entry = flow\_store\_cuckoo\_lkup(fid)}\;
\nl     \uIf(\tcp*[f]{\texttt{flow\_store} miss}){\textup{\texttt{entry} empty}}{
            \algstate{entry = flow\_store\_alloc()}\;
        }
\nl     \algstate{check\_memory\_safety(entry)}\;
\nl     \algstate{victim = drop\_from\_rear(flow\_cache)}\;
\nl     \algstate{victim = encrypt(victim)}\;
\nl     \algstate{swap(entry.state, victim.state)}\;
\nl     \algstate{entry = decrypt(entry)}\;
}
\nl     \algstate{raise\_to\_front(entry, flow\_cache)}\;
\nl \textbf{Return} \algstate{entry.state}\;
\caption{Fast flow tracking with dual lookup}
\label{alg:state}
\end{algorithm}

\subsection{Fast Flow Lookup}
The fastest path in the flow tracking process above is indicated by \texttt{flow\_cache} hit, where only a few pointers are updated to refresh LRU linkage. In case of \texttt{flow\_cache} miss and \texttt{flow\_store} hit, two memory copy (for swapping) and cryptographic operations are entailed. Due to the interlinked design, these operations have constant cost irrelevant to the number of tracked flows.

When encountering high flow concurrency, we found that the flow lookup sub-procedure becomes the main factor of performance slowdown, as confirmed by one of our tested middleboxes with an inefficient lookup design (PRADS, see Section~\ref{subsec:eval_mb}). 
Given the constrained enclave resources, two requirements are therefore imposed on the underlying lookup structure: \emph{search efficiency} and \emph{space efficiency.}

\inlsec{Dual lookup design with cuckoo hashing}
We recognize cuckoo hashing as the one to simultaneously achieve the two properties. It has guaranteed $O(1)$ lookup and superior space efficiency, e.g., $93\%$ load factor with two hash functions and a bucket size of $4$~\cite{a-cool-and-practical-alternative-to-traditional-hash-tables}. 
One downside with hashing is their inherent cache-unfriendiness~\cite{heileman2005caching}, which incurs a higher cache miss penalty in the enclave. Thus, while adopting cuckoo hashing, we still need a cache-aware design.

Our idea is to split \texttt{lkup\_table} into a small table dedicated for \texttt{flow\_cache}, and a much larger one for \texttt{flow\_store}. The large one is searched only after a miss in the small one.
The smaller table contains the same number of entries as \texttt{flow\_cache} and has a fixed size 
that can well fit into a typical L3 cache ($8$MB). It is accessed on every packet and thus is likely to reside in L3 cache most of the time. 
Such a dual lookup design can perform especially well when the \texttt{flow\_cache} miss rate is relatively low.

\begin{figure}
    \centering
    \begin{minipage}[t]{0.48\linewidth}
        \includegraphics[width=\linewidth]{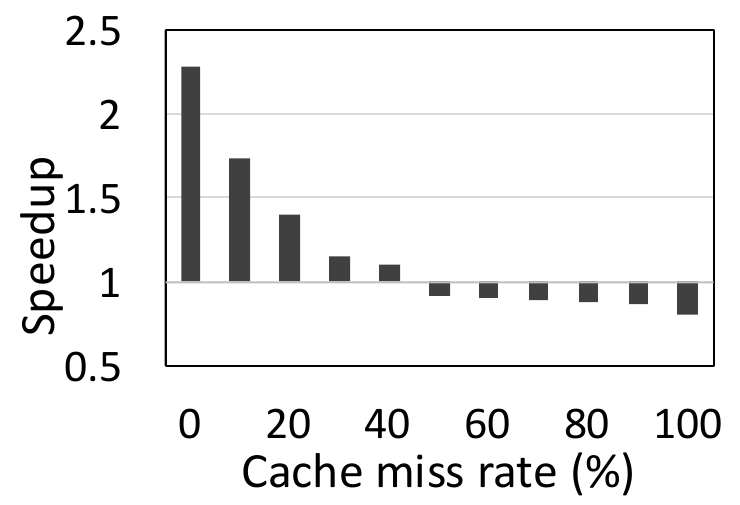}
        \caption{The speedup of dual lookup design over single lookup design.}
        \label{fig:dual_lkup}
    \end{minipage}
    \hfill
    \begin{minipage}[t]{0.48\linewidth}
        \includegraphics[width=\linewidth]{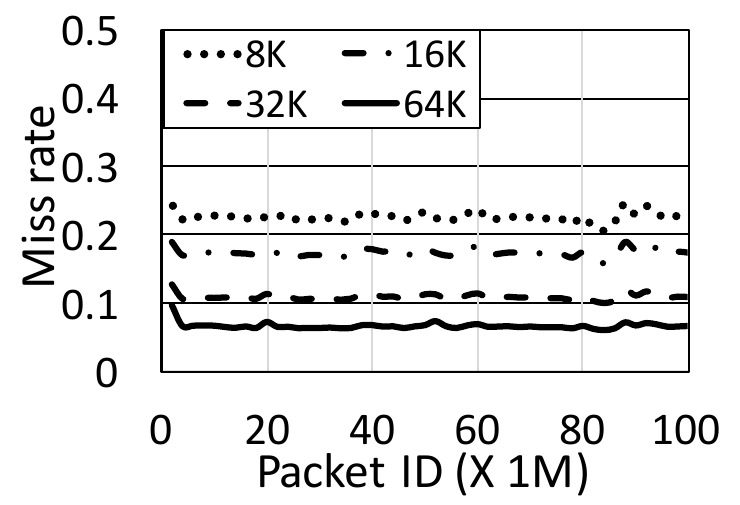}
        \caption{The miss rate of \texttt{flow\_cache} with varying size for a real network trace~\cite{caida}.}
        \label{fig:cache_miss}
    \end{minipage}
\end{figure}

\label{subsec:state_fast_lkup}
To validate the design, we evaluate the two lookup approaches with 1M flows, $512$B states and \texttt{flow\_cache} with $32$K entries. As expected, Figure~\ref{fig:dual_lkup} shows that the lower the miss rate, the larger speedup the dual lookup achieves over the single lookup. Real-world traffic often exhibits temporal locality~\cite{kim2009revisiting,casado2008}. We also estimate the miss rate of \texttt{flow\_cache} over a real trace~\cite{caida}. As shown in Fig.~\ref{fig:cache_miss}, the miss rate can be maintained well under $20\%$ with $16$K cache entries, confirming the temporal locality in the trace, hence the efficiency of the dual lookup design in practice.

\subsection{Security of State Management}
\label{subsec:state_security}
We show that the adversary can only gain little knowledge from the management procedures. It can neither manipulate the procedures to influence middlebox behavior. Therefore, the proposed management scheme retains the same security level as if it is not applied, i.e., when all states are handled by EPC paging.

We first analyze the adversary's view throughout the procedures. Among the three tables, \texttt{flow\_cache} and \texttt{lkup\_table} are always kept in the enclave, hence invisible to the adversary. Stored in untrusted memory, \texttt{flow\_store} is fully disclosed. The adversary can obtain all \texttt{store\_entry}'s, but never sees the state in clear text.

She will notice the creation of new flow state, but cannot link it to a previous one, even if the two have exactly the same content, because of the random initialization of the \texttt{swap\_count}. Similarly, she is not able to track traffic patterns (e.g., packets coming in bursts) of a single flow, because the \texttt{swap\_count} will increment upon each swapping and produce different ciphertexts for the same flow state. In general, she cannot link any two \texttt{store\_entry}'s.

The explicit termination of a flow is unknown to the adversary, as the procedure takes place entirely in the enclave. In contrast, she will notice state removal events during expiration checking. Yet, this information is useless as the entries are not linkable.

Now we consider an active adversary. Due to the authenticated encryption, any modification of \texttt{state\_entry}'s is detectable. Malicious deletion of a \texttt{state\_entry} will be also caught when it is supposed to be swapped into the enclave after a hit in \texttt{lkup\_table}. She cannot inject a fake entry since \texttt{lkup\_table} is inaccessible to her. Furthermore, the replay attack will be thwarted because \texttt{swap\_count} keeps the freshness of the state.

\section{Instantiations of \system}
\label{sec:impl_inst}
We have implemented a working prototype of \system~
and instantiated it for three case-study stateful middleboxes.

\subsection{Porting Middleboxes to SGX}
A middlebox system should be first ported to the SGX enclave before it can enjoy the security and performance benefits of \system, as illustrated in Fig.~\ref{fig:system_arch}. After that, the middlebox's original insecure I/O module will be seamlessly replaced with \texttt{etap} and the network frameworks stacked thereon; its flow state management procedures, including memory management, flow lookup and termination, will be changed to that of \system~as well.

There are several ways to port a legacy middlebox. 
One is to build the middlebox with trusted LibOS~\cite{tsai2017graphene,baumann2014}, which are pre-ported to SGX and support general system services within the enclave. Another more specialized approach is to identify only the necessary system services and customize a trusted shim layer for optimized performance and TCB size~\cite{lind2017glamdring}.
To prepare for our middlebox case-studies, we follow the second approach and implement a shim layer that supports the necessary system calls and \texttt{struct} definitions. 

Some prior systems allow modular development of middleboxes that are automatically secured by SGX~\cite{poddar2018safebricks,trach2018shieldbox,han2017}. For middleboxes built this way, we can directly substitute their network I/O and flow state management with \system, augmenting them with full-stack protection and efficient stateful processing.

\begin{figure*}
\begin{minipage}[t]{.32\linewidth}
    \centering
    \includegraphics[width=\linewidth]{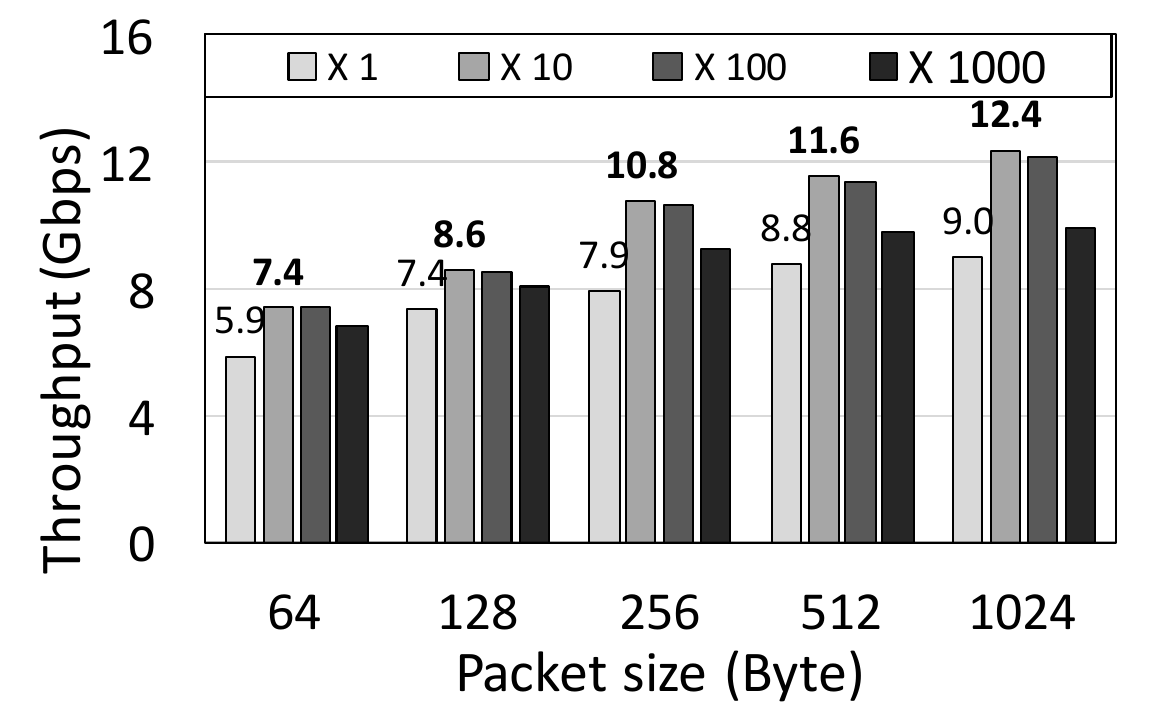}
    \caption{Performance of \texttt{etap} against varied batch size.}
    \label{fig:net_io_batching}
\end{minipage}  
\hfill
\begin{minipage}[t]{.32\linewidth}
    \centering
    \includegraphics[width=\linewidth]{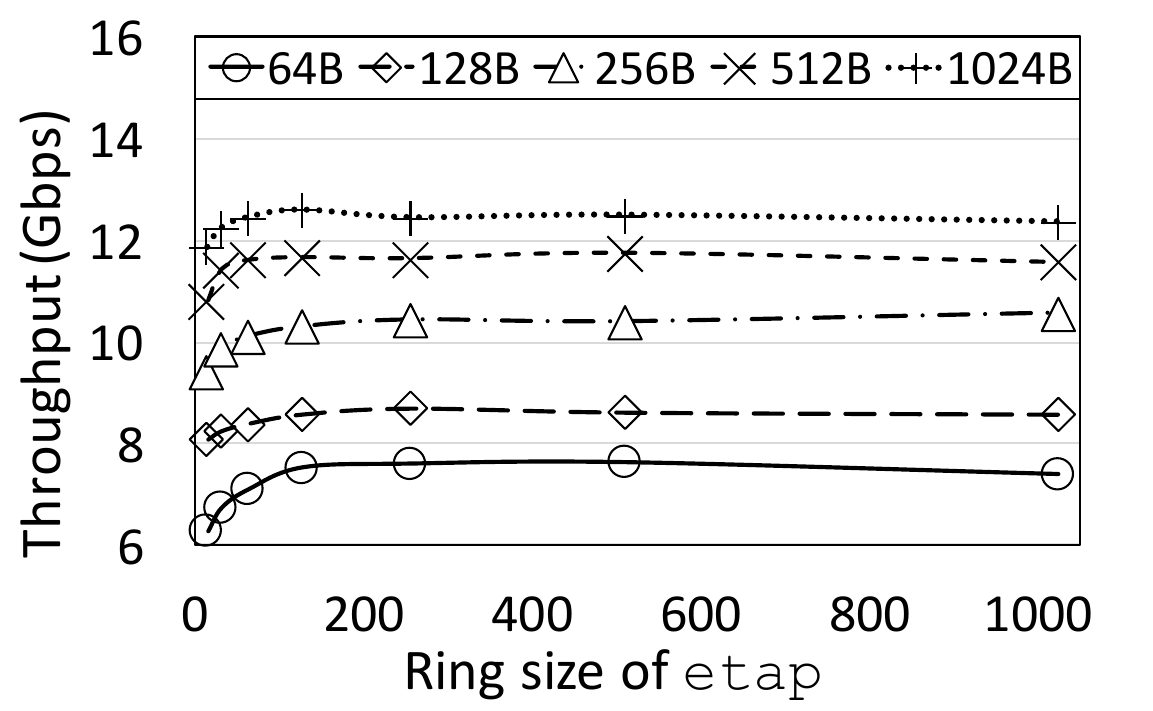}
    \caption{Performance of \texttt{etap} against varied ring size.}
    \label{fig:net_io_ring}
\end{minipage}
\hfill
\begin{minipage}[t]{.32\linewidth}
    \centering
    \includegraphics[width=\linewidth]{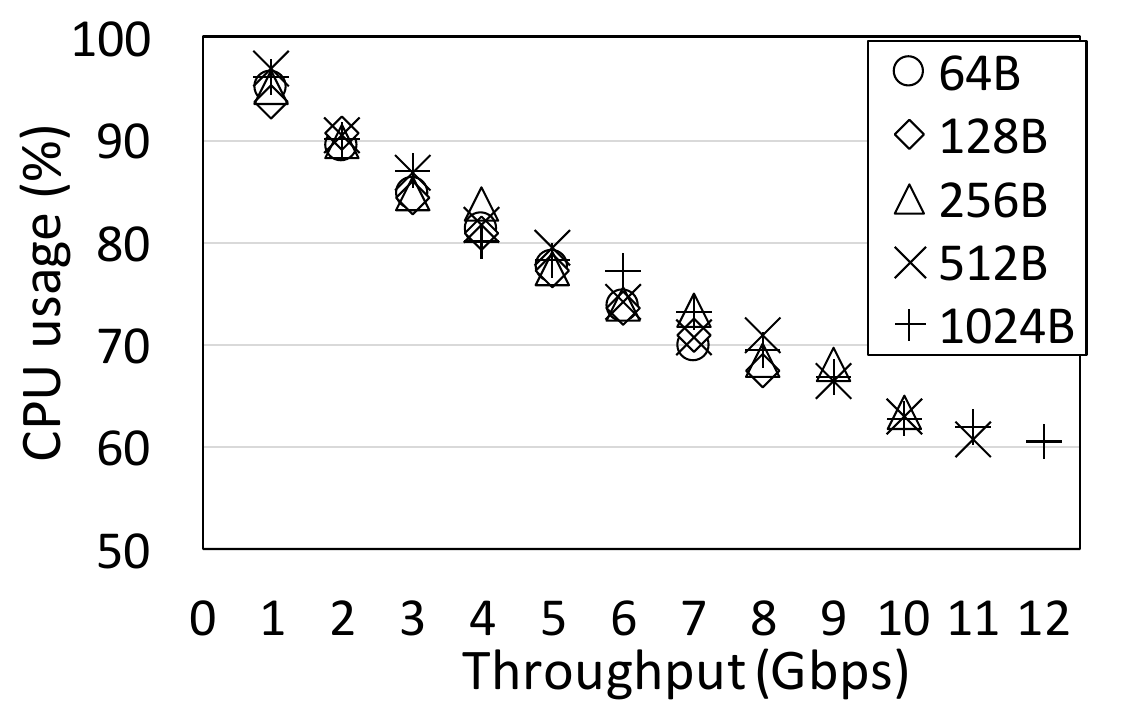}
    \caption{CPU usage of \texttt{etap} against middlebox throughput.}
    \label{fig:net_io_cpu}
\end{minipage}
\end{figure*}

\begin{figure}
    \centering
    \includegraphics[width=.9\linewidth]{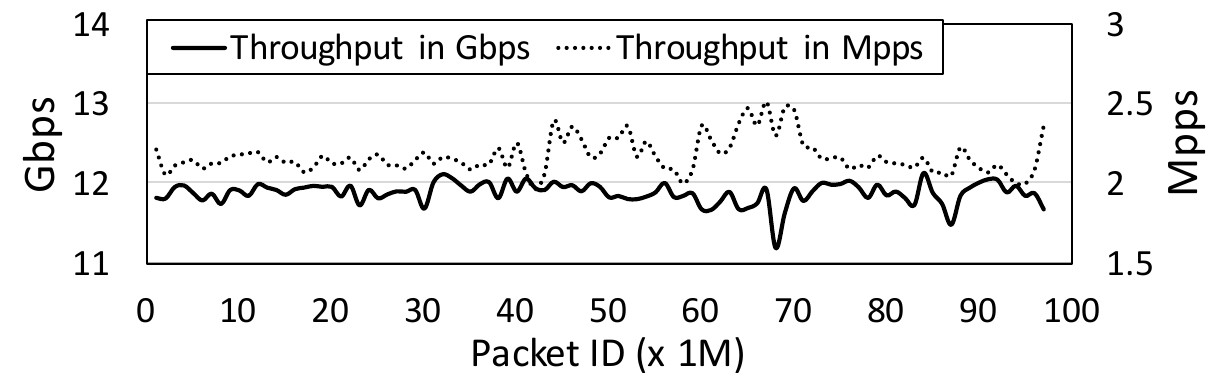}
    \caption{Performance of \texttt{etap} on real trace.}
    \label{fig:net_io_real}
\end{figure}

\subsection{Middlebox Case Studies}
We now introduce the three middleboxes we instantiated for \system{}. For discussions on the efforts of instantiating them with \system, we assume that they have already been ported to SGX. Both PRADS and lwIDS use \texttt{libpcre} for pattern matching, so we manually port it as a trusted library to be used within the enclave.

\inlsec{PRADS~\cite{PRADS}}
Capable of detecting network assets (e.g., OSes, devices) in packets against predefined fingerprints, PRADS has been widely used in academic research~\cite{jamshed17mos,khalid2016,Gember-Jacobson:2014:OEI:2619239.2626313}. 
It uses \texttt{libpcap} for packet I/O, so its main packet loop can be directly replaced with the compatibility layer we built on \texttt{etap} (Section~\ref{subsec:etap_use}). We also adapt its own flow tracking logic to \system's state management procedures without altering the original functionality. This affects about $200$ lines of code (LoC) in the original PRADS project with $10$K LoC.

\inlsec{lwIDS}
Based on the tcp reassembly library \texttt{libntoh}~\cite{libntoh}, we built a \underline{l}ight\underline{w}eight \underline{IDS} that can identify malicious patterns over reassembled data.
Whenever the stream buffer is full or the flow is completed, the buffered content will be flushed and inspected against a set of patterns. Note that the packet I/O and main stream reassembly logic of lwIDS is handled by \texttt{libntoh} (3.8K LoC), which we have already ported on top of \texttt{etap} (Section~\ref{subsec:etap_use}). The effort of instantiating \system~ for lwIDS thus reduces to adjusting the state management module of \texttt{libntoh}, which amounts to a change of around 100 LoC.

\inlsec{mIDS}
We design a more comprehensive middlebox, called mIDS, based on the mOS framework~\cite{jamshed17mos} and the pattern matching engine DFC~\cite{ChoiCJP16}. Similar to lwIDS, mIDS will flush stream buffers for inspection upon overflow and flow completion; but to avoid consistent failure, it will also do the flushing and inspection when receiving out-of-order packets, as we found that the logic for handling such case is yet to be completed in current mOS code. Again, since we have ported mOS (26K LoC) with \texttt{etap} (Section~\ref{subsec:etap_use}), the remaining effort of instantiating \system~ for mIDS is to modify the state management logic, resulting in $450$ LoC change. Note that such effort is one-time: hereafter, we can instantiate any middlebox built with mOS without change.

\section{Evaluation}
\label{sec:evaluation}
\begin{figure*}
\centering
\begin{minipage}[t]{.65\linewidth}
    \begin{minipage}{\linewidth}
        \begin{subfigure}{0.32\linewidth}
            \includegraphics[width=\linewidth]{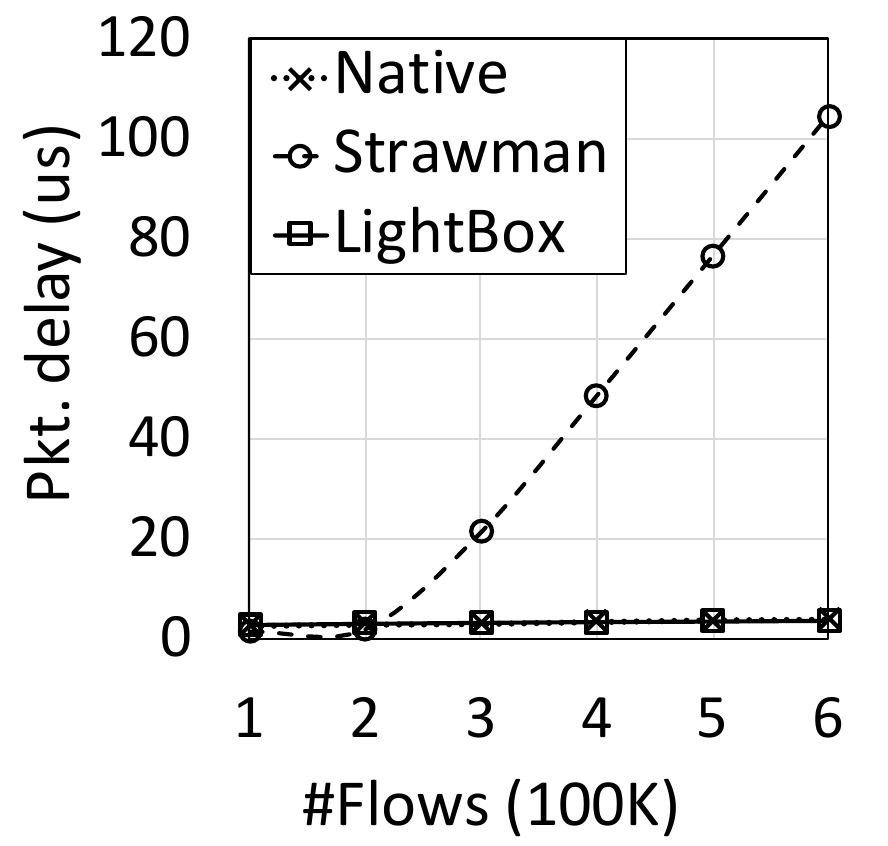}
            \caption{64B packet.}
            \label{fig:control_prads_64}
        \end{subfigure}
        \begin{subfigure}{0.32\linewidth}
            \includegraphics[width=\linewidth]{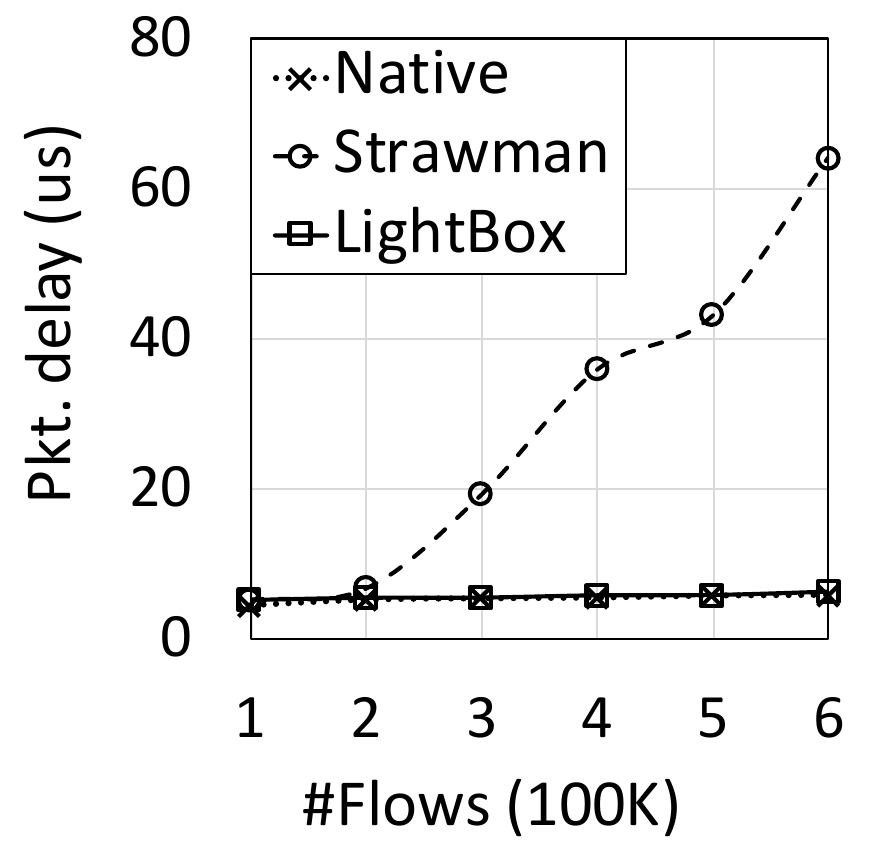}
            \caption{512B packet.}
            \label{fig:control_prads_512}
        \end{subfigure}
        \begin{subfigure}{0.32\linewidth}
            \includegraphics[width=\linewidth]{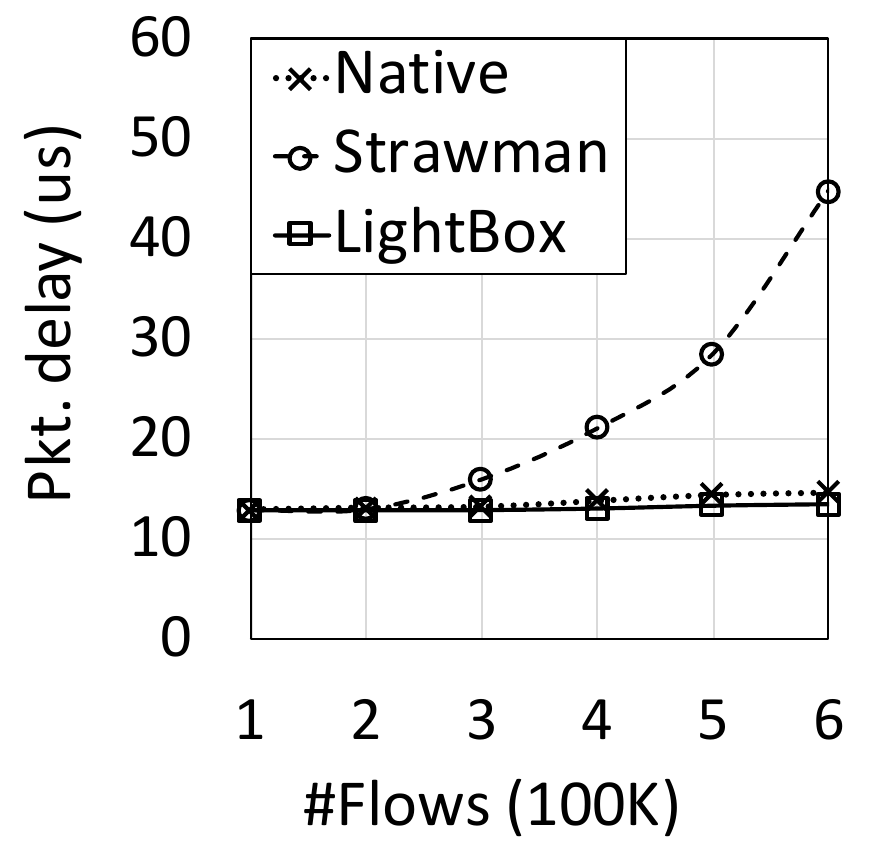}
            \caption{1500B packet.}
            \label{fig:control_prads_1500}
        \end{subfigure}
        \caption{Performance of PRADS under controlled settings.}
        \moresp{1}
        \label{fig:control_prads}
    \end{minipage}
    \begin{minipage}{\linewidth}
        \begin{subfigure}{0.32\linewidth}
            \centering
            \includegraphics[width=\linewidth]{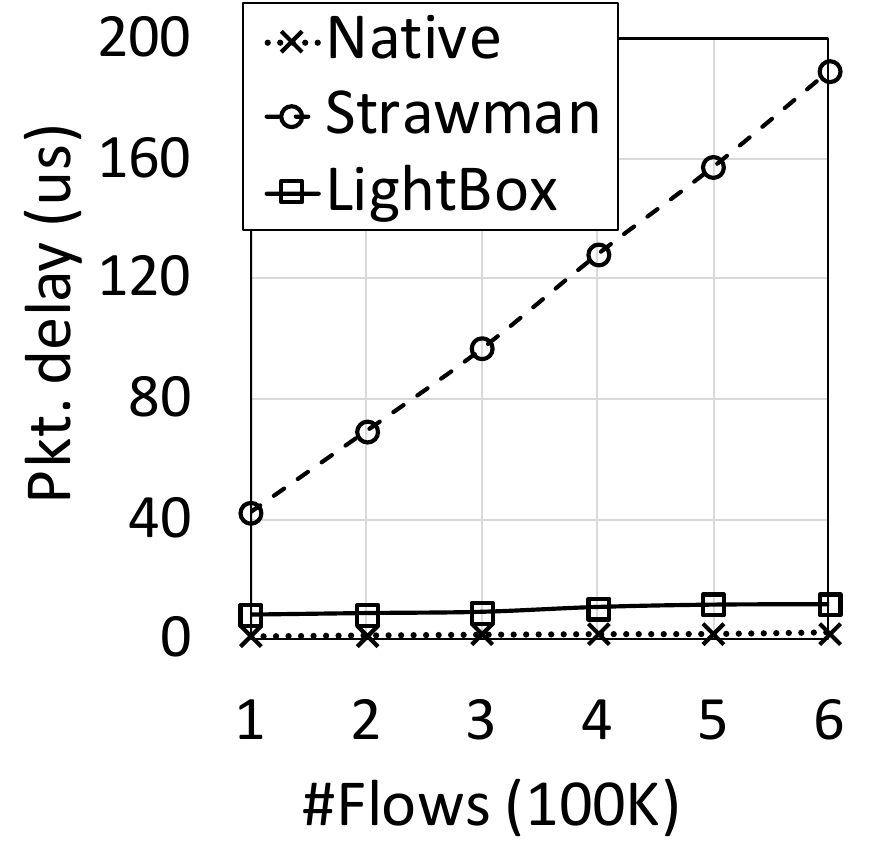}
            \caption{64B packet.}
            \label{fig:control_lwids_64}
        \end{subfigure}
        \begin{subfigure}{0.32\linewidth}
            \centering
            \includegraphics[width=\linewidth]{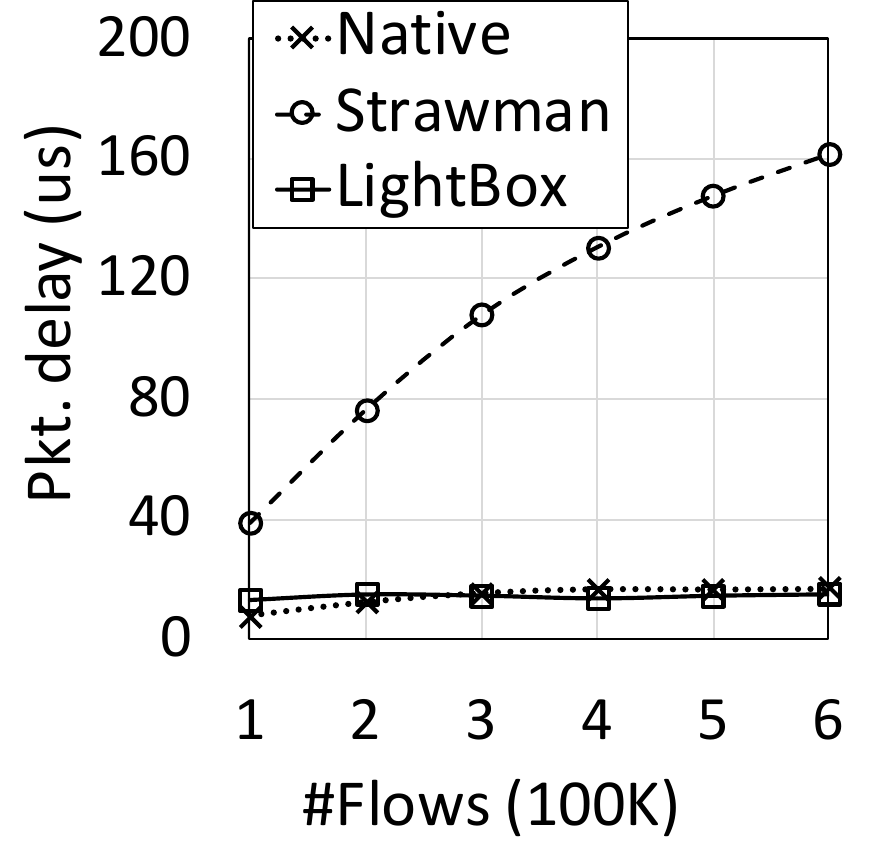}
            \caption{512B packet.}
            \label{fig:control_lwids_512}
        \end{subfigure}
        \begin{subfigure}{0.32\linewidth}
            \centering
            \includegraphics[width=\linewidth]{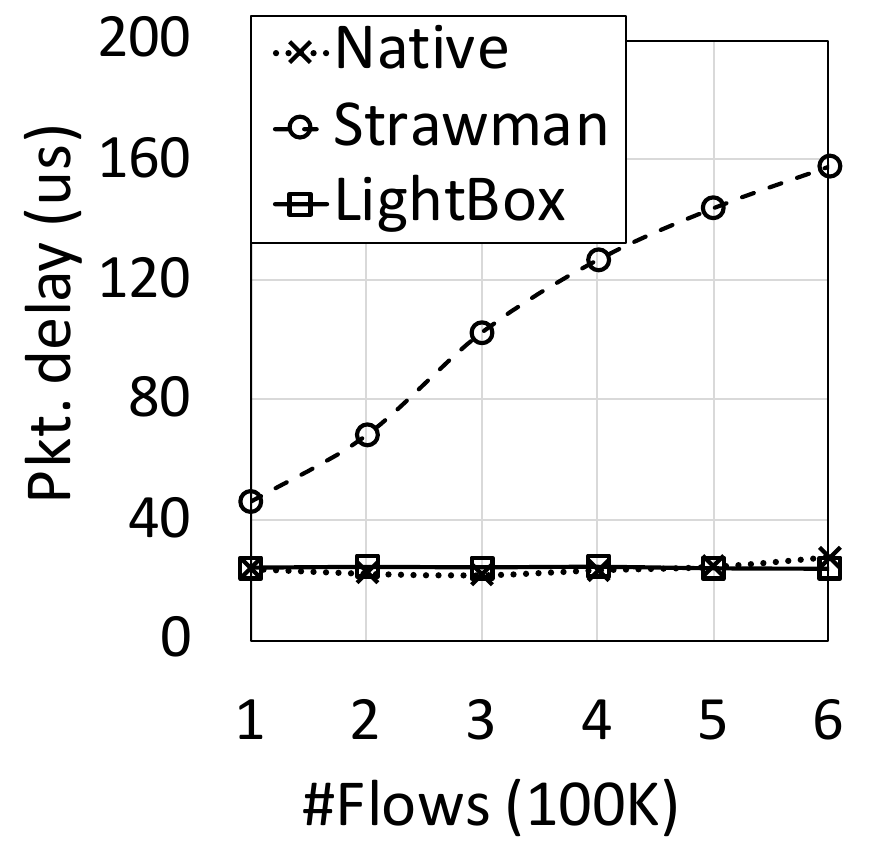}
            \caption{1500B packet.}
            \label{fig:control_lwids_1500}
        \end{subfigure}
        \caption{Performance of lwIDS under controlled settings.}
        \label{fig:control_lwids}
    \end{minipage}
\end{minipage}
\hfill
\begin{minipage}[t]{.34\linewidth}
    \begin{minipage}{\linewidth}
        \centering
        \includegraphics[width=\linewidth]{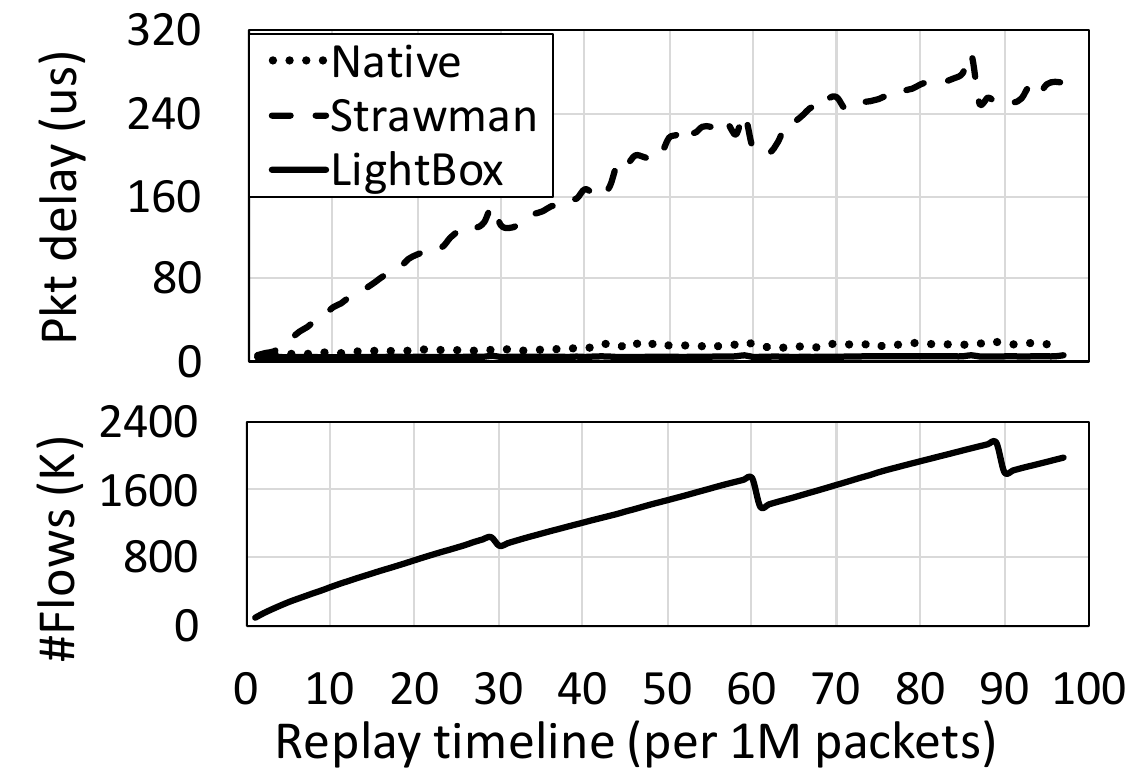}
        \caption{PRADS on real trace.}
        \moresp{1}
        \label{fig:real_prads}
    \end{minipage}
    \begin{minipage}{\linewidth}
        \includegraphics[width=\linewidth]{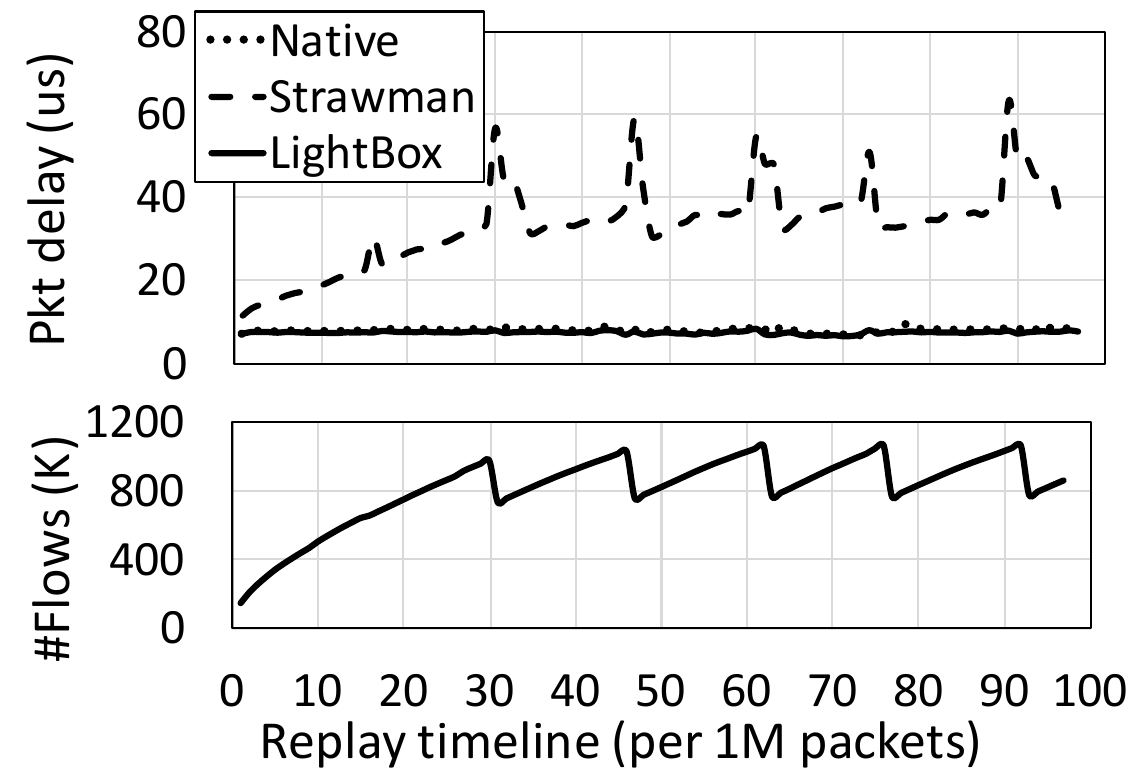}
        \caption{lwIDS on real trace.}
        \label{fig:real_lwids}
    \end{minipage}
\end{minipage}
\end{figure*}

\subsection{Methodology and Setup}
\label{subsec:eval_setup}
Our evaluation comprises two main parts: in-enclave packet I/O, where we evaluate \texttt{etap} from various aspects and decide the practically optimal configurations~(Section~\ref{subsec:eval_io}); middlebox performance, where we measure the efficiency of \system\ against a native and a strawman approach for the three case-study middleboxes~(Section~\ref{subsec:eval_mb}). We will also give discussions on experimental comparison between \system{} and previous systems (Section~\ref{subsec:exp_comp}). We use a real SGX-enabled workstation with Intel E3-1505 v5 CPU and 16GB memory in the experiments. Equipped with $1$Gbps NIC, the workstation is unfortunately incapable of reflecting \texttt{etap}'s real performance, so we prepare two experiment setups. In what follows, we will use K for thousand and M for million in the units.

\inlsec{Setup 1}
The first setup is dedicated for evaluation on \texttt{etap}, where we run \texttt{etap-cli} and \texttt{etap} on the same standalone machine and let them communicate with the fast memory channel via kernel networking. Note that \texttt{etap-cli} needs no SGX support and runs as a normal user-land program. To reduce the side effect of running them on the same machine, we tame the kernel networking buffers such that they are kept small ($500$KB) but still performant. Our intent here is to demonstrate that \texttt{etap} can catch up with the rate of a real 10Gbps NICs in practical settings.

\inlsec{Setup 2}
Deployed in a local 1Gbps LAN, the second setup is for evaluating middlebox performance. We use a separate machine as the \emph{gateway} to run \texttt{etap-cli}, so it communicates with \texttt{etap} via the real link. The gateway machine also serves as the \emph{server} to accept connections from clients (on other machines in the LAN). We then use \texttt{tcpkali}~\cite{tcpkali} to generate concurrent TCP connections transmitting random payloads from clients to the server; all ACK packets from the server to clients are filtered out. Our environment can afford up to $600$K concurrent connections. We also obtain a real trace from CAIDA~\cite{caida} for experiments; it is collected by monitors deployed at backbone networks. The trace is sanitized and contains only anonymized L3/L4 headers, so we pad them with random payloads to their original lengths specified in the header. We use the first $100$M packets from the trace in our experiments. 

\subsection{In-enclave Packet I/O Performance}
\label{subsec:eval_io}
To evaluate \texttt{etap}, we create a bare middlebox which keeps reading packets from \texttt{etap} without further processing. It is referred to as PktReader. 
We keep a large memory pool ($8$GB) and feed packets to \texttt{etap-cli} directly from the pool. 

\inlsec{Parameterized evaluation}
We first investigate how batching size affects \texttt{etap} performance. The ring size is set as $1024$. As shown in Fig.~\ref{fig:net_io_batching}, the optimal size appears between $10$ and $100$ for all packet sizes. The throughput drops when the batching size becomes either too small or overly large, matching our expectation and analysis in Section~\ref{subsec:etap_perf}. With a batching size of $10$, \texttt{etap} can deliver small $64$B (byte) packet at $7.4$Gbps, and large $1024$B packet at $12.4$Gbps, which is comparable to advanced packet I/O framework on modern $10$Gbps NIC~\cite{Rizzo2012}. We set $10$ as the default batching size and use this configuration in all following experiments.

Shrinking \texttt{etap} ring is beneficial in that precious enclave resources can be saved for middlebox functions, and in the case of multi-threaded middleboxes, for efficiently supporting more RX rings. However, smaller ring size generally leads to lower I/O throughput. Figure~\ref{fig:net_io_ring} reports the results with varying ring sizes. As can be seen, the tipping point occurs around $256$, where the throughput for all packet sizes begins to drop sharply as ring size decreases. Beyond that and up to $1024$, the performance appears insensitive to ring size. We thus use $256$ as the default ring size in all subsequent tests. 

\inlsec{Resource consumption}
The rings contribute to the major \texttt{etap} enclave memory consumption. One ring uses as small as $0.38$MB as per the default configuration, and a working \texttt{etap} consumes merely $0.76$MB.
The core driver of \texttt{etap} is run by dedicated threads and we are also interested in its CPU consumption. The driver will spin in the enclave if the rings are not available, since exiting enclave and sleeping outside is too costly. This implies that a slower middlebox thread will force the core driver to \emph{waste} more CPU cycles in the enclave. To verify such effect, we tune PkgReader with different levels of complexity, and estimate the core driver's CPU usage under varying middlebox speed. As expected, the results in Fig.~\ref{fig:net_io_cpu} delineate a clear negative correlation between the CPU usage of \texttt{etap} and the performance of middlebox itself. With $70\%$ utilization of a single core the core driver can handle packets at its full speed. Overall, we can see that an average commodity processor is more than enough for our target 10Gpbs in-enclave packet I/O.

\inlsec{Performance on real trace}
Figure~\ref{fig:net_io_real} shows \texttt{etap}'s performance on the real CAIDA trace that has a mean packet size of 680B. We estimate the throughput for every 1M packets while replaying the trace to \texttt{etap-cli}. As shown, although there are small fluctuations overtime due to varying packet size, the throughput remains mostly within $11-12$Gbps and $2-2.5$Mpps. This further demonstrates \texttt{etap}'s practical I/O performance.

\begin{figure*}
\centering
\begin{minipage}{.65\linewidth}
  \begin{subfigure}{0.32\linewidth}
      \includegraphics[width=\linewidth]{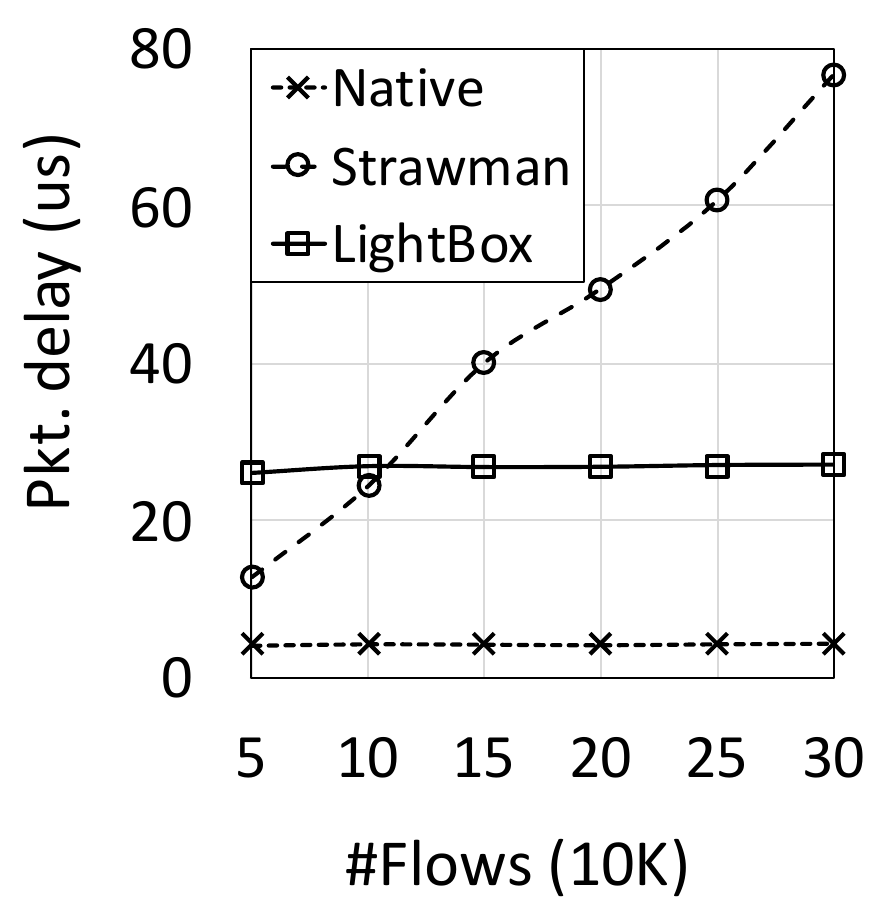}
      \caption{64B packet.}
      \label{fig:control_mids_64}
  \end{subfigure}
  \begin{subfigure}{0.32\linewidth}
      \includegraphics[width=\linewidth]{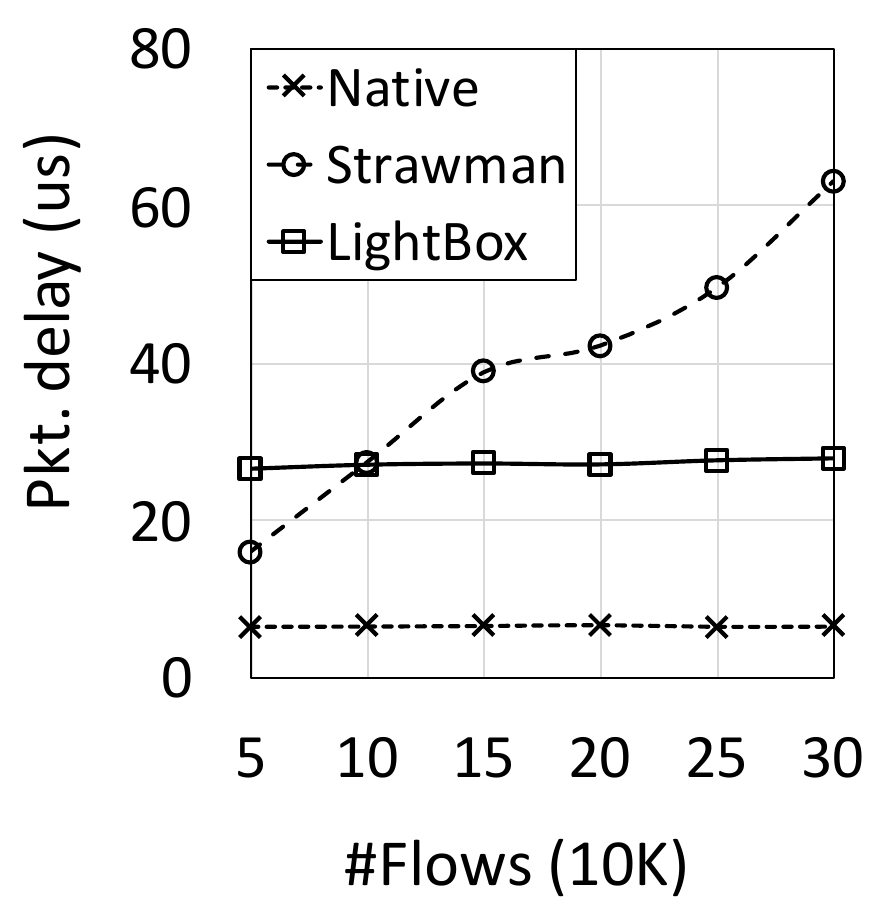}
      \caption{512B packet.}
      \label{fig:control_mids_512}
  \end{subfigure}
  \begin{subfigure}{0.32\linewidth}
      \includegraphics[width=\linewidth]{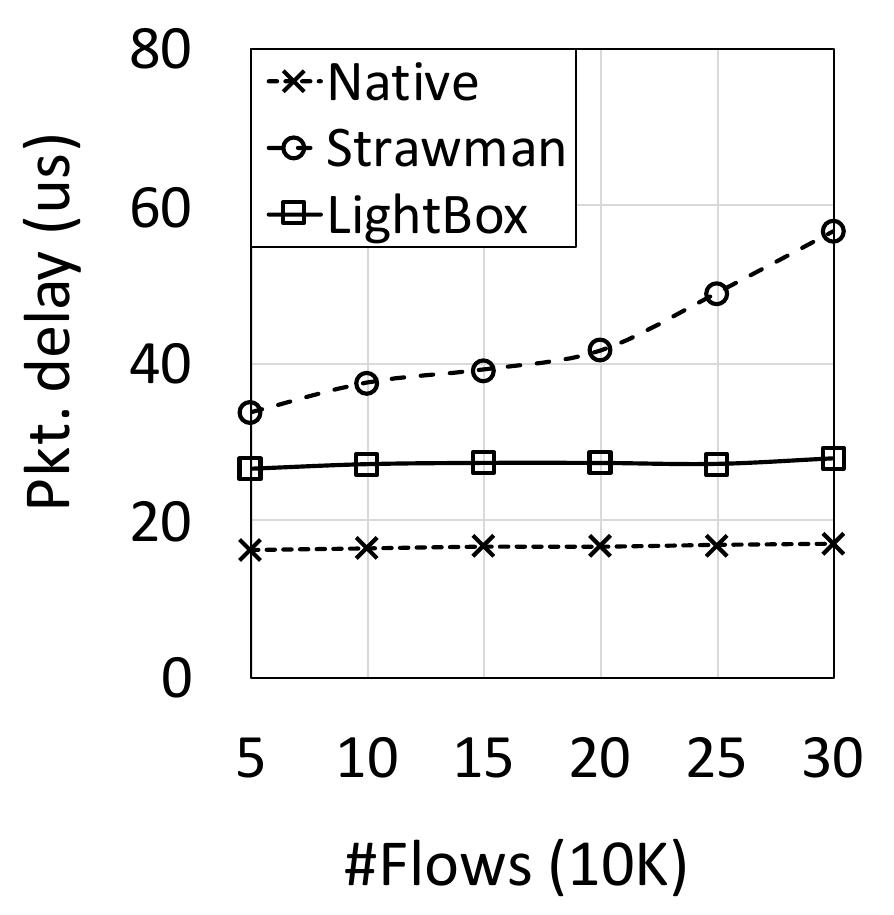}
      \caption{1500B packet.}
      \label{fig:control_mids_1500}
  \end{subfigure}
  \caption{Performance of mIDS under controlled settings.}
  \label{fig:control_mids}
\end{minipage}
\hfill
\begin{minipage}{.34\linewidth}
  \includegraphics[width=\linewidth]{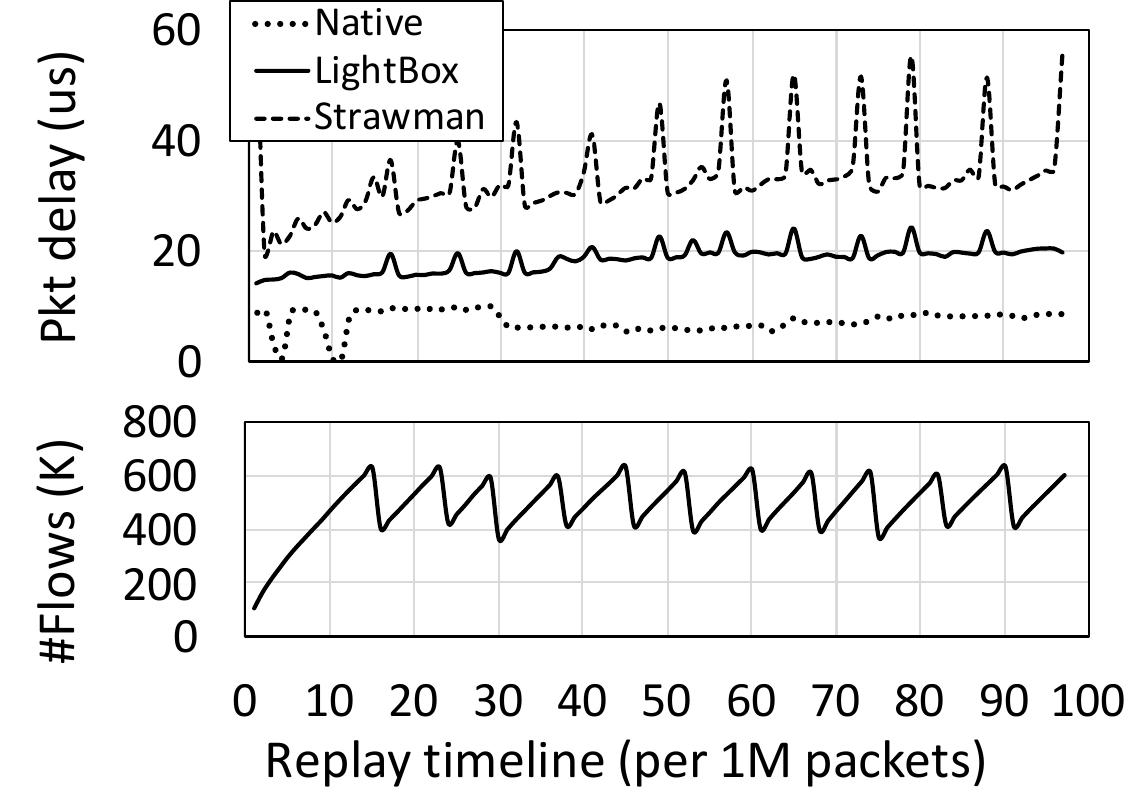}
  \caption{mIDS on real trace.}
  \label{fig:real_mids}
\end{minipage}
\end{figure*}

\subsection{Middlebox Performance}
\label{subsec:eval_mb}
We study the performance of the three middleboxes, each with three variants: the vanilla version (denoted as Native) running as a normal program;  naive SGX port (denoted as Strawman) that uses \texttt{etap} and our ported \texttt{libntoh} and mOS for networking, but relies on EPC paging for however much enclave memory is needed; the \system\ instance as described in Section~\ref{sec:impl_inst}. It is worth noting that despite the name, the Strawman variants actually benefit a lot from \texttt{etap}'s efficiency. Our goal here is primarily to investigate the efficiency of our state management design.

We use the default configurations for all three middleboxes unless otherwise specified. For lwIDS we compile $10$ \texttt{pcre} engines with random patterns for inspection; for mIDS we build the DFC engine with $3700$ patterns extracted from Snort community ruleset. The flow state of PRADS, lwIDS, and mIDS has a size of $512$B\footnote{PRADS has $124$B flow state, which is too small under our current experiment settings. To better approximate realistic scenarios, we pad the flow state of PRADS to $512$B with random bytes. No such padding is applied to lwIDS and mIDS.}, $5.5$KB, and $11.4$KB\footnote{This size is resulted from the rearrangement of mOS's data structures pertaining to flow state. We merge all data structures into a single one to ease memory management.}, respectively; the latter two include stream reassembly buffer of size $4$KB and $8$KB.
For LightBox variants, the number of entries of \texttt{flow\_cache} is fixed to $32$K, $8$K and $4$K for PRADS, lwIDS, and mIDS, respectively.

\subsubsection{Controlled live traffic}
To gain a better understanding of how stateful middleboxes behave in the highly constrained enclave space, we test them in controlled settings with varying number of concurrent TCP connections between clients and the server. 
We control the clients' traffic generation load such that the aggregated traffic rate at the server side remains roughly the same for different degrees of concurrency. By doing so the comparisons are made fair and meaningful. In addition, we start to collect data points only when all connections are established and stabilized. We measure the mean packet processing delay in microsecond (\us) every 1M packets, and each reported data point is averaged over $100$ runs.

\inlsec{PRADS}
From Fig.~\ref{fig:control_prads}, we can see that \system\ adds negligible overhead ($<1$\us) to native processing of PRADS regardless of the number of flows. In contrast, Strawman incurs significant and increasing overhead after $200$K flows, due to the involvement of EPC paging. Interestingly, by comparing the subfigures it can also be seen that Strawman performs worse for smaller packets. This is because smaller packet leads to higher packet rate while saturating the link, which in turn implies higher page fault ratio. For $600$K flows, \system\ attains $3.5\times$ --- $30\times$ speedup over the Strawman.

\inlsec{lwIDS}
Figure~\ref{fig:control_lwids} presents similar results for lwIDS. Here, the performance of Strawman is further degraded, since lwIDS has larger flow state size than PRADS and its memory footprint exceeds $550$MB even when tracking only $100$K flows. For $64$B packet, \system\ introduces $6-8$\us~ packet delay ($4-5\times$ to native) because the state management dominates the whole processing; nonetheless, it still outperforms Strawman by $5-16\times$. For larger packets, the network function itself becomes dominant and the overhead of \system\ over Native is reduced, as shown in Fig.~\ref{fig:control_lwids} (b) and (c).

\inlsec{mIDS}
Among the case-study middleboxes, mIDS is the most complicated one with the largest flow state. Here, our testbeds can scale to $300$K concurrent connections. For each connection mIDS will track two flows, one for a direction, and allocate memory accordingly. But since we filter out the trivial ACK packets from the server to clients, we still count only one flow per connection. 
Figure~\ref{fig:control_mids} reveals that the performance of mIDS's three variants follows similar trends as in previous middleboxes: Native and \system{} are insensitive to the number of concurrent flows; conversely, the overhead of Strawman grows as more flows are tracked.

But in contrast to previous cases, now the overhead of \system{} over Native becomes notable. This is explained by mIDS's large flow state size, i.e., $11.4$ KB, which leads to the substantial cost of encrypting/decrypting and copying states. Besides, we found that for each packet, in addition to its own flow, mIDS will also access the paired flow, \emph{doubling the cost of our flow tracking design} (see Section~\ref{subsec:state_mgmt}). Nonetheless, we can see that the gap is closing towards larger packet size, as the network function processing itself weighs in. Later in this section we will discuss how to further improve our design to cope with large flow state and connection-based tracking.

\subsubsection{Real trace}
Now we investigate middlebox performance with respect to the real CAIDA trace. The trace is loaded by the gateway and replayed to the middlebox for processing. Again, we collect the data points every $1$M packets. Packets of unsupported types are filtered out so only $97$ data points are collected for each case. Since L2 headers are stripped in the CAIDA trace, we also adjust the packet parsing logic accordingly for the middleboxes. Yet another important factor for real trace is the flow timeout setting. We must carefully set the timeout so inactive flows are purged well in time, lest excessive flows overwhelm the testbeds. Here, we set the timeout for PRADS, lwIDS, and mIDS, to $60$, $30$, and $15$ seconds, respectively. Table~\ref{tbl:mb_caida} reports the overall throughout of relaying the trace. Below we give more detailed analysis.

\inlsec{PRADS}
Figure~\ref{fig:real_prads} shows that the packet delay of Strawman grows with the number of flows; it needs about $240$\us~to process a packet when there are $1.6$M flows. In comparison, \system\ maintains low and stable delay (around $6$\us) throughout the test. A bit surprisingly, it even edges over the native processing as more flows are tracked, attributed to an inefficient chained hashing design used in the native implementation. This highlights the importance of \emph{efficient flow lookup in stateful middleboxes}.

\inlsec{lwIDS}
Compared with PRADS, the number of concurrent flows tracked by lwIDS decreases, as shown in Fig.~\ref{fig:real_lwids}. This is due to the halved timeout and the more aggressive strategy we used for flow deletion: we remove a flow when a \texttt{FIN} or \texttt{RST} flag is received, and we do not handle \texttt{TIME\_WAIT} event. It can be seen that with fewer flows, Strawman still incurs remarkable overhead, while the difference between \system~ and Native is indistinguishable.

\inlsec{mIDS}
The case for mIDS is tricky. Its current implementation of flow timeout seems not to be fully working, so we replaced the related code with the logic of checking all flows for expiration every timeout interval. We also made some modifications to ensure that the packet formats and abnormal packets in the real trace can be properly processed. Figure~\ref{fig:real_mids} reports the test results. There is again a large gap between Strawman and Native. Yet, as in the controlled settings, there is some moderate gap between \system{} and Native, due to the large state and double flow tracking design.

\begin{table}
    \centering
    \caption{Throughput (Mbps) under CAIDA trace.}
    \begin{tabular}{@{}cccc@{}}
        \toprule
        & Native & Strawman & LightBox \\ \midrule
        \multicolumn{1}{l|}{PRADS} &    429.24   &  67.399   &     928.06     \\
        \multicolumn{1}{l|}{lwIDS} &    689.11    &   182.57  &      685.36    \\ 
        \multicolumn{1}{l|}{mIDS} &    713.56    &   161.02  &      310.42    \\ \bottomrule
    \end{tabular}
    \label{tbl:mb_caida}    
\end{table}

\subsubsection{Reflections on future improvement}
Above results show that when the per-flow state size is not overly large, our current design, which treats the state as a whole chunk of raw data in an agnostic manner, suffices to achieve near-native performance for stateful processing. Otherwise, it may not be wise to manage the large state as a whole. We would expect more fine-grained partition and handling of the state to improve efficiency.

A promising direction is to separate the large stream buffer (e.g., $8$KB in the case of mIDS) from the rest of the state (e.g., $3.4$KB). The processing of each packet will only touch a small portion of the buffer, dispensing with the high overhead of encrypting, decrypting and copying the entire buffer. Only when it is time to flush the buffer for inspection do we need to load it into enclave in its entirety. This will significantly diminish the cost of the flow tracking routine. Note that here we should refrain from moving encrypted raw packets individually out from the enclave, as this will leak the packet size and count. A more secure way would be to divide the stream buffer into chunks of fixed length and handle packets in batches.

In a similar vein, for middleboxes that access the flow state of both connection directions on each packet, instead of treating both flows equally, we can manage only the necessary data fields of the paired flow, leaving the vast majority of its state untouched. This requires a slight redesign of our current data structures to support the effective linking of pairing flows.

A common theme in the fine-grained approaches suggested above is to reduce the amount of unnecessary data moved across the enclave boundary, and hence lessen the management overhead. We leave the detailed designs and exploration of potential trade-offs between security and efficiency as an interesting future work.

\subsection{Comparison with Previous Systems}
\label{subsec:exp_comp}
We now discuss previous secure middlebox systems built upon SGX regarding experiment settings and performance evaluation. They have been evaluated on various network functions, and they demonstrate that certain workloads can be run in the enclave with marginal performance overhead. This work differs from them in that it takes into consideration: 1) the complexity of stateful processing as seen in production-level middleboxes, and 2) the high flow concurrency encountered in deployed networks. These two features together pose unique challenges in operating stateful middleboxes in enclaves at a reasonable cost. To our best knowledge, such experiment settings have not been considered in prior works. Their experiments are mostly confined to small memory footprint within EPC limit ($128$MB), avoiding expensive EPC paging; but in our settings, the memory footprint of middleboxes can grow to multiple GBs. We discuss some representative systems below.

In ShieldBox~\cite{trach2018shieldbox}, several stateless middleboxes, including some micro ones with simplistic functions, are evaluated. They process packets independently, without tracking any flow states. Since the middlebox memory footprint is always kept small, it is not very surprising to see that the performance of shielded middleboxes is close to native in virtually all test cases.

SGX-BOX~\cite{han2017} allows inspection over reassembled streams. However, it leaves flow state other than the stream buffer outside enclave unprotected, obviating the challenge of fully protecting stateful middlebox processing. Besides, it reports only preliminary evaluation that the overhead of inspecting a single stream is small.

Safebricks~\cite{poddar2018safebricks} presents a more diverse set of experiments. But the middleboxes under testing are still arguably too simple in reality. Unlike the stateful lwIDS and more advanced mIDS we have built, it uses a simple DPI application working on individual packets but not reassembled streams. The NAT and load balancer tested there are also stateless. It does include a stateful firewall, but the functionality seems basic, and more importantly, it was not evaluated against concurrent flows.

To sum up, due to the different focuses and methodology, it is not feasible to derive a direct and fair experimental comparison between \system{} and previous systems. We for the first time evaluate non-trivial stateful middleboxes under settings with high flow concurrency, and we hope that this work can invite more efforts on bridging the research-practice gap for secure middlebox systems. On the other hand, in addition to our intensive evaluations, it is also interesting and critical to experiment with other important aspects such as service function chaining, as done by previous works~\cite{trach2018shieldbox,poddar2018safebricks}. Our current work has built solid a foundation for conducting further evaluations, and we leave them as future work.

\section{Related Work}
\label{sec:related}
\inlsec{Secure Middleboxes}
BlindBox~\cite{SherryCRS15} is the first system that applies cryptographic protocols (i.e., searchable encryption and garbled circuit) to enable inspection on encrypted packet payloads. A list of follow-up designs are proposed. The work~\cite{YuanWLW16} emphasizes on the protection of middlebox rules and the support of more inspection rules. SPABox~\cite{fan2017spabox} and BlindIDS~\cite{canard2017blindids} put extra attention on reducing session setup cost. Besides payloads, privacy-preserving packet header checking is also studied. The work~\cite{MelisAC15} considers using heavy homomorphic encryption for generic network functions, which is mostly of theoretical interest only. SplitBox~\cite{AsgharMSDK16} employs a distributed model for a certain class of packet header processing with multi-party computation techniques. Embark~\cite{LanSP16} introduces a customized prefix-matching scheme; by integrating the technique from~\cite{SherryCRS15}, it supports a wider class of middlebox functions. These software-centric solutions are often restricted in functionality (especially for stateful processing) and performance.

There are also several designs based on trusted hardware, i.e., Intel SGX. S-NFV~\cite{shih2016} proposes to protect specifically the middlebox state, but not the entire middlebox processing over protected traffic. Trusted Click~\cite{coughlin2017} and ShieldBox\cite{trach2018shieldbox} port Click modular router~\cite{kohler2000click} to the enclave. But they do not consider the protection of metadata, and lack support for stateful processing due to the limitation of Click that is inherently stateless. In a setting different from middlebox outsourcing, SGX-BOX~\cite{han2017} and mbTLS~\cite{Naylor2017} enable middleboxes to intercept TLS connections and securely inspect traffic in the enclave, with primary focus on programmability and deployability, respectively. One note on SGX-BOX is that it employs the mOS framework for stateful processing outside enclave, so all states except the encrypted stream buffers are left unprotected and the correctness of stateful processing is not guaranteed. Targeting the same scenario as us, the latest work SafeBricks~\cite{poddar2018safebricks} pays extra attention to middlebox code protection, and applies IPSec for secure traffic tunneling. While protecting packet headers, it is still vulnerable to traffic analysis attacks leveraging packet size and count. None of these hardware-assisted solutions protects low-level traffic metadata as \system\ does, nor do they enable efficient stateful processing in real networks experiencing high flow concurrency.

\inlsec{Reducing overhead of SGX}
One theme of our research is to minimize the performance overhead incurred by SGX while retaining security guarantees. A main strand of works approach this goal with switch-less \texttt{ECALLs}/\texttt{OCALLs} by avoiding expensive context switching~\cite{weisse2017,arnautov2016}. This approach is orthogonal to our designs. In fact, with the official release of such support~\cite{Tian2018SCM}, we can readily replace our normal use of \texttt{OCALL} with the switch-less version for even better efficiency. Eleos~\cite{orenbach2017} introduces user-managed paging to alleviate the overhead of naive EPC paging. Its cache-store architecture is similar to ours, but for generality it entails coarse-grained data structure (e.g., page table) and complex procedures (e.g,. address translation and page table walk), compared with our tailored and optimized designs. Therefore, it may not suit the performance-sensitive middlebox applications. To help grow EPC in the future, a recent study proposes to refine the underlying data structure for integrity checking~\cite{Taassori2018VRP}. Whether a large EPC can overcome SGX's current performance issue without enlarging the attack surface remains an open problem. Our state management design significantly increases the nominal secure memory that is usable by SGX applications. We hope that the proposed compact data structure and efficient lookup algorithm can provide helpful insights into future study towards this direction.

\inlsec{SGX-enabled systems} 
Many systems have been built with SGX, such as data analytics platforms~\cite{schuster2015, Shaon2017} and secure system services~\cite{arnautov2016, hunt2016}. Their application scenarios are different from ours, and in particular, they do not face the challenge of operating stateful middlebox in real networks with strict performance requirements.

\section{Conclusion}
\label{sec:conclusion}
We present \system, an SGX-assisted secure middlebox system. While many researches have explored the possibilities of securing middleboxes with SGX and claimed its efficiency and practicality, this is the first work to ascertain the claims from a more pragmatic perspective. We bridge the research-practice gap by identifying two critical challenges, from both security and functionality aspect, and satisfactorily address them with domain knowledge and extensive customization. Among others, our first main technical contribution is an elegant in-enclave virtual network interface that is highly secure, efficient and usable; and our second main innovation is the flow state management scheme comprising data structures and algorithms optimized for the enclave space. They together build up a comprehensive solution for deploying off-site middleboxes with strong protection and stateful processing, at near-native speed. 
We hope that \system\ can push forward secure network function virtualization and middlebox-as-a-service to the practical realm, and that our work can invite more efforts on building hardware-assisted secure systems that are practical and usable.

%
\bibliographystyle{ACM-Reference-Format}
\bibliography{Lightbox-Arxiv-v3}

%
\appendix

\section{Multi-threading Support}
\label{sec:multi_threading}
\inlsec{RSS Emulation for \texttt{etap}}
\label{subsubsec:rss}
Many middleboxes utilize multi-threading to achieve high throughput~\cite{jamshed2012kargus,jamshed17mos,SekarER12,Dobrescu:2009:REP:1629575.1629578}. The standard parallel architecture used by them relies on receiver-side scaling (RSS) or equivalent software approaches to distribute traffic into multiple queues by flows. Each flow is processed in its entirety by one single thread without affecting the others. We equip \texttt{etap} with an emulation of this NIC feature to cater for multi-threaded middleboxes.

With the emulation, multiple RX rings will be created by \texttt{etap}, and each middlebox thread is binded to one RX ring. The core driver will hash the 5-tuple to decide which ring to push a packet, and the poll driver will only read packets from the ring binded to the calling thread. As the number of rings increases, the size of each ring should be kept small to avoid excessive enclave memory consumption. Note that we have discussed practical ring size in Section~\ref{sec:evaluation}.

\inlsec{Multi-threaded State Management}
The RSS mechanism ensures that each flow is processed in isolation to others. For a multi-threaded middlebox, we assign each thread a separate set of \texttt{flow\_cache}, \texttt{lkup\_table}, and \texttt{flow\_store}. There is no intersection between the sets, and thus all threads can perform flow tracking simultaneously without data racing. Note that compared to the single-threaded case, this partition scheme does not change memory usage in managing the same number of flows.

\section{Extension of Service Model}
\label{sec:extension}
To clearly lay out the core designs of \system, so far we have focused on a basic service model. That is, a single middlebox, and a single service provider hosting the middlebox service. Now we discuss how some other typical scenarios can be readily supported.

\begin{figure}
    \centering
    \includegraphics[width=.85\linewidth]{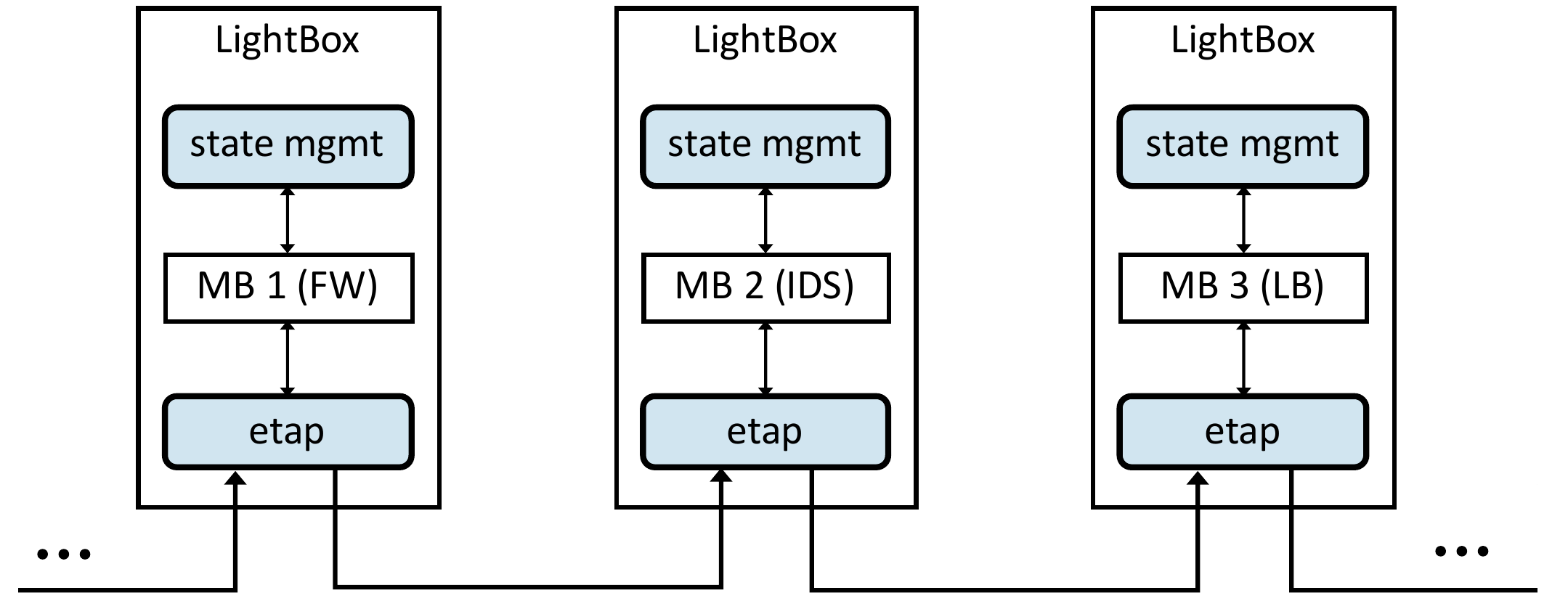}
    \caption{Service function chain connected by \texttt{etap}'s.}
    \label{fig:chain}
\end{figure}

\inlsec{Service function chaining}
Sometimes multiple logical middleboxes are chained together to process network traffic, which is commonly referred to as service function chaining~\cite{jain2016,sun2017nfp}. This service model is also considered in two recent systems for secure middlebox outsourcing. SafeBricks~\cite{poddar2018safebricks} chains the middleboxes within the same enclave, and isolates them by enforcing least privilege on each. In comparison, ShieldBox~\cite{trach2018shieldbox} chains the middleboxes with different enclaves run by different processes on the same physical machine. Both designs, however, run the chain on a single physical machine that has a limited EPC size, and do not consider the resource-demanding stateful middlebox. Practical execution of a single stateful middlebox in the enclave is already a non-trivial task --- what we strive to achieve in this paper --- let alone running multiple enclaved stateful middleboxes on the same machine, where severe performance issue is almost inevitable.

To this end, we propose to drive each middlebox in the chain with a \system\ instance on a separate physical machine.
Along the chain, one instance's \texttt{etap} will be simultaneously peered with previous and next instance's \texttt{etap} (or the \texttt{etap-cli} at the gateway). Now each \texttt{etap}'s core driver will effectively forward the encrypted traffic stream to the next \texttt{etap}. This way, each middlebox in the chain can access packet at line rate and run at its full speed. Note that the secure bootstrapping should be adjusted accordingly. In particular, the network administrator needs to attest each \system, and provision it with proper peer information. 

\inlsec{Disjoint service providers}
Middlebox outsourcing may span a disjoint set of service providers. A primary one may provide the networking and computing platform, yet others (e.g., professional cybersecurity companies) can provide bespoke middlebox functions and/or processing rules. Such service market segmentation calls for finer control over the composition of the security services.

The SGX attestation utility enables any participant of the joint service to attest enclaves on the primary service provider's platform. Therefore, they can securely provision their proprietary code/rule set to a trusted bootstrapping enclave. The code is then compiled in the bootstrapping enclave, and together with the rules, provisioned to \system\ enclave. 
Such on-the-fly compilation of private code in the enclave is first described in~\cite{schuster2015}. In~\cite{poddar2018safebricks}, it is applied to bootstrap secure middleboxes with different network function vendors. 
We refer interested readers to~\cite{poddar2018safebricks} for more details.

\end{document}